\documentclass[aps,prd,          groupedaddress,amsmath,amssymb]{revtex4}
\usepackage{graphicx}
\usepackage{dcolumn}
\usepackage{bm}

\begin{document}

\def\d{{\rm d}}
\def\lp{\left. }
\def\rp{\right. }
\def\lr{\left( }
\def\rr{\right) }
\def\le{\left[ }
\def\re{\right] }
\def\lg{\left\{ }
\def\rg{\right\} }
\def\lb{\left| }
\def\rb{\right| }
\def\beq{\begin{equation}}
\def\eeq{\end{equation}}
\def\bea{\begin{eqnarray}}
\def\eea{\end{eqnarray}}

\preprint{KA-TP-02-2007}
\preprint{LPSC 07-002}
\preprint{hep-ph/0701202}
\title{Threshold Resummation for Slepton-Pair Production at Hadron
 Colliders}
\author{Giuseppe Bozzi}
\affiliation{Institut f\"ur Theoretische Physik, Universit\"at
  Karlsruhe, Postfach 6980, D-76128 Karlsruhe, Germany}
\author{Benjamin Fuks}
\author{Michael Klasen}
\email[]{klasen@lpsc.in2p3.fr}
\affiliation{Laboratoire de Physique Subatomique et de Cosmologie,
 Universit\'e Joseph Fourier/CNRS-IN2P3,
 53 Avenue des Martyrs, F-38026 Grenoble, France}
\date{\today}
\begin{abstract}
We present a first and extensive study of threshold resummation effects for
supersymmetric (SUSY) particle production at hadron colliders, focusing on
Drell-Yan like slepton-pair and slepton-sneutrino associated production.
After confirming the known next-to-leading order (NLO) QCD corrections and
generalizing the NLO SUSY-QCD corrections to the case of mixing squarks in
the virtual loop contributions, we employ the usual Mellin $N$-space
resummation formalism with the minimal prescription for the inverse
Mellin-transform and improve it by resumming $1/N$-suppressed and a class
of $N$-independent universal contributions. Numerically, our results
increase the theoretical cross sections by 5 to 15\% with respect to the NLO
predictions and stabilize them by reducing the scale dependence from up to
20\% at NLO to less than 10\% with threshold resummation.
\end{abstract}
\pacs{12.60.Jv,13.85.Ni,14.80.Ly}
\maketitle


\section{Introduction}
\label{sec:1}

The Minimal Supersymmetric Standard Model (MSSM) \cite{Nilles:1983ge,%
Haber:1984rc} is one of the most promising extensions of the Standard Model
(SM) of particle physics. It postulates a symmetry between fermionic and
bosonic degrees of freedom in nature, predicting thus the existence of a
fermionic (bosonic) supersymmetric (SUSY) partner for each bosonic
(fermionic) SM particle. Its main advantages are the stabilization of the
gap between the Planck and the electroweak scale \cite{Witten:1981nf}, gauge
coupling unification at high energy scales \cite{Dimopoulos:1981yj}, and a
stable lightest supersymmetric particle as a dark matter candidate
\cite{Ellis:1983ew}. Spin partners of the SM particles have not yet been
observed, and in order to remain a viable solution to the hierarchy problem,
SUSY must be broken at low energy via soft mass terms in the Lagrangian. As
a consequence, the SUSY particles must be massive in comparison to their SM
counterparts, and the Tevatron and the LHC will perform a conclusive search
covering a wide range of masses up to the TeV scale.

Scalar leptons are among the lightest supersymmetric particles in many
SUSY-breaking scenarios \cite{Aguilar-Saavedra:2005pw} and often decay into
the corresponding Standard Model (SM) partner and the lightest stable SUSY
particle. A possible signal for slepton-pair production at hadron colliders
would thus consist in a highly energetic lepton pair and associated missing
energy. An accurate calculation of the transverse-momentum spectrum
\cite{Bozzi:2006fw} allows us to use the Cambridge (s)transverse mass to
measure the slepton masses \cite{Lester:1999tx} and spin \cite{Barr:2005dz}
and to distinguish this signal from the SM background, which is mainly due
to $WW$ and $t \bar t$ production \cite{Lytken:22,Andreev:2004qq}. Current
experimental (lower) limits on electron, muon, and tau slepton masses are 73
GeV, 94 GeV, and 81.9 GeV, respectively \cite{Yao:2006px}. The leading-order
(LO) cross section for the production of non-mixing slepton-pairs has been
calculated in \cite{Dawson:1983fw,Chiappetta:1985ku,delAguila:1990yw,%
Baer:1993ew}, while the mixing between the interaction eigenstates was
included in \cite{Bozzi:2004qq}. The next-to-leading order (NLO) QCD
corrections have been calculated in \cite{Baer:1997nh}, and the full
SUSY-QCD corrections have been added in \cite{Beenakker:1999xh}. However,
the presence of massive non-mixing squarks and gluinos in the loops makes
the genuine SUSY corrections considerably smaller than the standard QCD
ones.

In this paper, we extend this last work by including mixing effects relevant
for the squarks appearing in the loops, and we consider the
threshold-enhanced contributions, due to soft-gluon emission from the
initial state. These contributions arise when the initial partons have just
enough energy to produce the slepton pair in the final state. In this case,
the mismatch between virtual corrections and phase-space suppressed
real-gluon emission leads to the appearance of large logarithmic terms
$\alpha_s^n[\ln^{2n-1}(1-z)/(1-z)]_+$ at the $n^{{\rm th}}$ order of
perturbation theory, where $z=M^2/s$, $M$ is the slepton-pair invariant
mass, and $s$ is the partonic center-of-mass energy. When $s$ is close to
$M^{2}$, the large logarithms have to be resummed, i.e.\ taken into account
to all orders in $\alpha_{s}$. The convolution of the partonic cross section
with the steeply falling parton distributions enhances the threshold
contributions even if the hadronic threshold is far from being reached,
i.e.\ $\tau=M^2/S \ll 1$, where $S$ is the hadronic center-of-mass energy.
Large corrections are thus expected for the Drell-Yan production of a
slepton pair with invariant mass $M$ of a few 100 GeV at the Tevatron and
LHC.

All-order resummation is achieved through the exponentiation of the
soft-gluon radiation, which does not take place in $z$-space directly, but
in Mellin $N$-space, where $N$ is the Mellin-variable conjugate to $z$ and
the threshold region $z\rightarrow 1$ corresponds to the limit $N\rightarrow
\infty$. Thus, a final inverse Mellin-transform is needed in order to obtain
a resummed cross section in $z$-space. Threshold resummation for the
Drell-Yan process was first performed in \cite{Sterman:1986aj,Catani:1989ne}
at the leading-logarithmic (LL) and next-to-leading-logarithmic (NLL)
levels, corresponding to terms of the form $\alpha_{s}^{n}\ln^{2n}N$ and
$\alpha_{s}^{n}\ln^{2n-1}N$. The extension to the NNLL level
($\alpha_{s}^{n}\ln^{2n-2}N$ terms) has been carried out both for the
Drell-Yan process \cite{Vogt:2000ci} and for Higgs-boson production
\cite{Catani:2003zt}. Very recently, even the NNNLL contributions
($\alpha_{s}^{n}\ln^{2n-3}N$ terms) became available \cite{Moch:2005ba,%
Moch:2005ky,Laenen:2005uz}.

In this paper, we will perform resummation for slepton-pair production at
the NLL level, the reason being that away from the threshold region, the
resummed calculation has to be matched to the fixed-order calculation, which
is at present only known to NLO accuracy. Our analytical results for neutral
($\gamma$, $Z^0$) and charged ($W^\pm$) current slepton-pair and
slepton-sneutrino associated production at NLO will be presented in Sec.\
\ref{sec:2}. In Sec.\ \ref{sec:3}, our fixed-order results will be used to
perform the threshold-resummation to NLL accuracy, and numerical predictions
will be made in Sec.\ \ref{sec:4}. We summarize our results in Sec.\
\ref{sec:5}. The mixing of sfermion interaction eigenstates is discussed in
App.\ \ref{sec:a}, and the NLO SUSY-QCD form factors for mixing squark
loop contributions are collected in App.\ \ref{sec:b}.

\section{Slepton-Pair Production at Fixed Order in Perturbative QCD}
\label{sec:2}

In this section, we present the leading and next-to-leading order
contributions in the strong coupling constant $\alpha_s$ to the
mass-spectrum for slepton-pair and slepton-sneutrino associated production
in hadronic collisions through neutral- and charged-current Drell-Yan type
processes,
\bea
\label{eq:proc}
 h_a(p_a)\,h_b(p_b)&\to&\tilde{l}_i(p_1)\,\tilde{l}^{(\prime)\ast}_j(p_2).
\eea
We define the square of the weak coupling constant $g_W^2=e^2/\sin^2
\theta_W$ in terms of the electromagnetic fine structure constant
$\alpha=e^2/(4\pi)$ and the squared sine of the electroweak mixing angle
$x_W=\sin^2\theta_W$. The coupling strengths of left- and right-handed
(s)fermions to the neutral and charged electroweak currents are then given
by
\bea
\label{eq:smZ}
 \{ L_{Z f f}, R_{Z f f} \} &=& 2\,T^{3}_f - 2\,e_f  \,x_W, \\
\label{eq:susyZ}
 \{ L_{Z \tilde{f}_i \tilde{f}_j}, R_{Z \tilde{f}_i \tilde{f}_j} \} &=&
 \{ L_{Z f f}\, S^{\tilde{f}}_{i1}\, S^{\tilde{f}\ast}_{j1}, R_{Z f f}\,
 S^{\tilde{f}}_{i2}\, S^{\tilde{f}\ast}_{j2}\, \},\\
\label{eq:smW}
 L_{W f f^\prime} &=& \sqrt{2}\,\cos\theta_W\,V_{f f^\prime}, \\
\label{eq:susyW}
 L_{W \tilde{f}_i \tilde{f}^\prime_j} &=& L_{W f f^\prime}\,
 S^{\tilde{f}}_{i1}\, S^{\tilde{f}^\prime\ast}_{j1},
\eea
where the weak isospin quantum numbers are $T_f^3=\pm1/2$ for left-handed
and $T_f^3=0$ for right-handed (s)fermions, their fractional electromagnetic
charges are denoted by $e_f$, and $V_{f f^\prime}$ are the usual CKM-matrix
elements. In general SUSY-breaking models, the sfermion interaction
eigenstates are not identical to the respective mass eigenstates, and mixing
effects must be included in the coupling strengths through the unitary
matrices $S^{\tilde{f}}$ diagonalizing the sfermion mass matrices (see App.\
\ref{sec:a}). For purely left-handed sneutrino eigenstates $\tilde{\nu}_L$ a
diagonalizing matrix is not needed, i.e.\ $S^{\tilde{\nu}}_{L1}$=1 and
$S^{\tilde{\nu}}_{ij}$=0 otherwise. In non-minimal flavour violating (NMFV)
models with inter-generational mixing, the squark and slepton (sneutrino)
mass matrices become in principle six- (three-) dimensional, although
flavour-changing neutral lepton currents are experimentally strongly
constrained \cite{Bartl:2005yy}.

Thanks to the QCD factorization theorem, the unpolarized hadronic cross
section
\bea
\label{eq:cross1}
 \sigma = \sum_{a,b}\int_{\tau}^1 {\rm d}x_a \int_{\tau/x_a}^1 {\rm d}x_b\,
 f_{a/h_a}(x_a,\mu_F^2)\, \ f_{b/h_b}(x_b,\mu_F^2) \,\hat{\sigma}_{ab}
 \lr z, M^2;\alpha_s(\mu_R^2),\frac{M^2}{\mu_F^2}, \frac{M^2}{\mu_R^2}\rr
\eea
can be written as the convolution of the relevant partonic cross section
$\hat{\sigma}_{ab}$ with the universal distribution functions $f_{a,b/
h_{a,b}}$ of partons $a,b$ inside the hadrons $h_{a,b}$, which depend
on the longitudinal momentum fractions of the two partons $x_{a,b}$ and
on the unphysical factorization scale $\mu_F$. The partonic scattering cross
section will be expressed in terms of the SUSY particle masses
$m_{\tilde{l}_i}$, $m_{\tilde{\nu}}$, $m_{\tilde{q}_i}$, $m_{\tilde{g}}$ and
the masses of the neutral and charged electroweak gauge bosons $m_Z$ and
$m_W$. The dependence on the strong coupling constant $\alpha_s$, the
factorization and renormalization scales $\mu_F$ and $\mu_R$, the invariant
mass of the slepton pair $M$ and the scaling variable $z=M^2/s$, where
$s=x_ax_bS$ and $S=(p_a+p_b)^2$, will be explicitly shown. In QCD
perturbation theory, the partonic cross section can be expanded in powers of
$\alpha_s$,
\bea
\label{eq:cross}
 \hat{\sigma}_{ab} \lr z, M^2; \alpha_s(\mu_R^2), \frac{M^2}{\mu_F^2},
 \frac{M^2}{\mu_R^2}\rr &=& \sum_{n=0}^\infty\left(\frac{\alpha_s(\mu_R^2)}
 {\pi}\right)^n \sigma_{ab}^{(n)}\lr z,M^2; \frac{M^2}{\mu_F^2}, \frac{M^2}
 {\mu_R^2}\rr.
\eea
In this work we have computed the LO ($n=0$) and NLO ($n=1$) coefficients in
the case of general mixing between the sfermion interaction eigenstates.

\subsection{Total Cross Sections at Leading Order}
\label{sec:2a}

\begin{figure}
\centering
\includegraphics[width=.3\columnwidth]{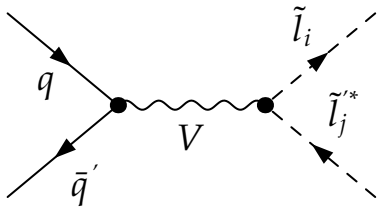}
\caption{\label{fig:1}Feynman diagram for slepton-pair ($V=\gamma, Z^0$) and
 slepton-sneutrino associated ($V=W^\mp$) production at leading order in
 perturbative QCD.}
\end{figure}

At leading order in perturbative QCD, slepton-pair and associated
slepton-sneutrino production proceed through an $s$-channel exchange of a
photon, a $Z^0$- or a $W^\mp$-boson,
\bea
\begin{array}
 {l}q\bar{q} \to \gamma, Z^0 \to \tilde{l}_i\,\tilde{l}^\ast_j,\\
 q\bar{q}^\prime \to W^\mp \to \tilde{l}_i\,\tilde{\nu}_l^\ast,\,
 \tilde{l}^\ast_i \,\tilde{\nu}_l,
\end{array}
\eea
as shown in Fig.\ \ref{fig:1}. The corresponding cross sections were first
%
%
calculated in \cite{Dawson:1983fw,Chiappetta:1985ku,delAguila:1990yw,%
Baer:1993ew}, and mixing effects between the interaction eigenstates
relevant for third-generation sleptons were added in \cite{Bozzi:2004qq}.

For neutral currents, the first coefficient of Eq.\ (\ref{eq:cross}) is
given by
\bea
\label{eq:sig0n}
 \sigma_{q\bar{q}}^{(0)}\lr z,M^2;\frac{M^2}{\mu_F^2},\frac{M^2}{\mu_R^2}\rr
 &=& \sigma_0(M^2)\,\delta (1-z)\nonumber\\
 &=& \frac{\alpha^2\, \pi\, \beta^3}{9\,M^2} \Bigg[e_q^2\, e_l^2\,
 \delta_{ij} + \frac{e_q\,e_l\,\delta_{ij} (L_{Z q q} + R_{Z q q})\,
 {\rm Re}\left[L_{Z \tilde{l}_i\tilde{l}_j} + R_{Z \tilde{l}_i
 \tilde{l}_j}\right]}{4\, x_W\,(1-x_W)\, (1-m_Z^2/M^2)} \nonumber\\
 && \hspace*{12mm}+\,\frac{(L_{Z q q}^2 + R_{Z q q}^2) \left|L_{Z
 \tilde{l}_i\tilde{l}_j} + R_{Z \tilde{l}_i \tilde{l}_j}\right|^2}
 {32\,x_W^2\, (1-x_W)^2 (1-m_Z^2/M^2)^2}\Bigg]\, \delta(1-z).
\eea
The three terms in Eq.\ (\ref{eq:sig0n}) represent the squared
photon-contribution, the photon-$Z^0$ interference and the squared
$Z^0$-contribution, respectively. The slepton-mass dependence is
factorized in the velocity
\bea
 \beta &=& \sqrt{1 + m_i^4/M^4 + m_j^4/M^4 - 2(m_i^2/M^2 + m_j^2/M^2 +
 m_i^2\,m_j^2/M^4)}.
\eea

The purely left-handed charged-current cross section is easily
derived from Eq.\ (\ref{eq:sig0n}) by setting
\bea
 m_Z \to m_W,~~~
 e_q = e_l = R_{Z q q} = R_{Z \tilde{l}_i \tilde{l}_j} = 0, ~~~
 L_{Z q q} \to L_{W q q^\prime} {\rm ,~~and~~}
 L_{Z \tilde{f}_i \tilde{f}_j} \to L_{W \tilde{f}_i \tilde{f}_j^\prime},
\eea
which gives
\bea
 \sigma_{q\bar{q}^\prime}^{(0)}\lr z, M^2;\frac{M^2}{\mu_F^2},\frac{M^2}
 {\mu_R^2}\rr &=& \sigma_0^\prime(M^2)\, \delta(1-z)\nonumber\\
 &=& \frac{\alpha^2\, \pi\, \beta^3}{9\, M^2} \Bigg[\frac{\left|
 L_{W q q^\prime}L_{W \tilde{l}_i \tilde{\nu}_l} \right|^2}{32\, x_W^2\,
 (1-x_W)^2 (1-m_W^2/M^2)^2}\Bigg]\, \delta(1-z).
\eea

After integration of the differential cross sections presented in
\cite{Baer:1993ew,Bozzi:2004qq}, we find agreement with the neutral-current
result of \cite{Dawson:1983fw} and the charged-current result of
\cite{Baer:1993ew} in the limit of non-mixing mass-degenerate sleptons,
obtained by setting all mixing matrices to the identity and by summing over
the left- and right-handed eigenstates, as well as with the general charged-
and neutral-current results of \cite{Bozzi:2004qq}.

\subsection{Next-to-Leading Order SUSY-QCD Corrections with Mixing Squark Loops}
\label{sec:2b}

\begin{figure}
\centering
\includegraphics[width=.9\columnwidth]{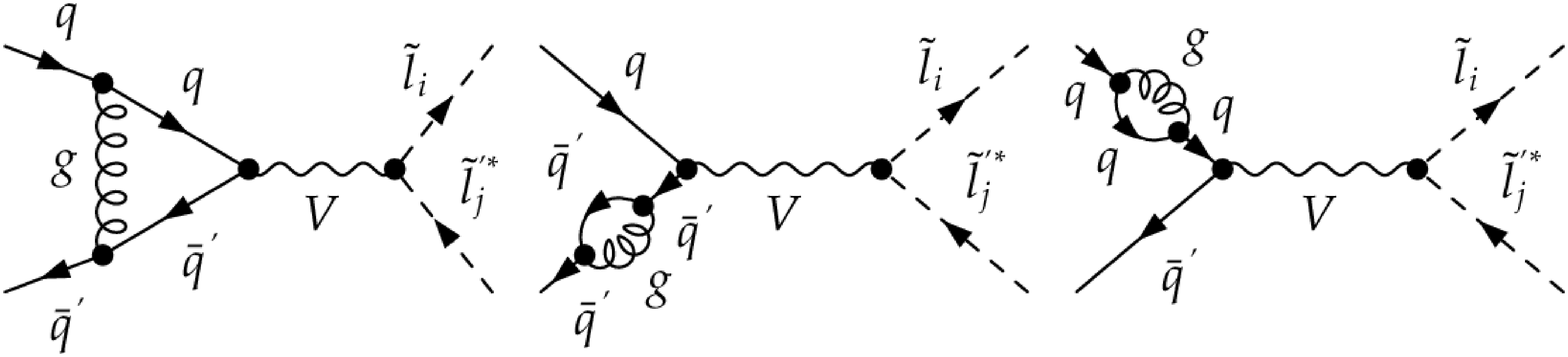}
\includegraphics[width=.9\columnwidth]{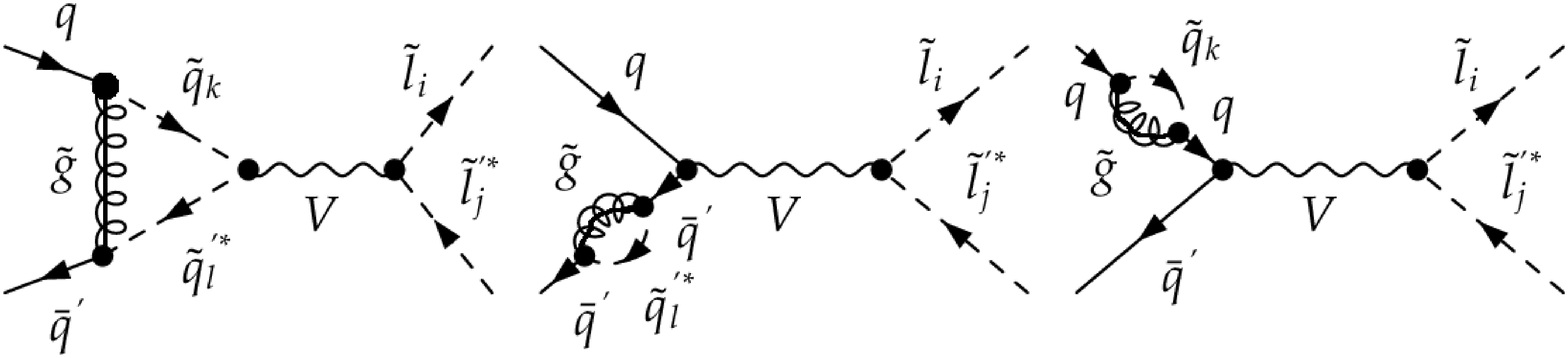}
\caption{\label{fig:2}Contributions of virtual diagrams for slepton-pair
 ($V=\gamma, Z^0$) and slepton-sneutrino associated ($V=W^\mp$) production
 at next-to-leading order in perturbative QCD. The first and second lines
 show the QCD and SUSY-QCD corrections, respectively. In the SUSY-QCD case,
 one has to sum over squark mass-eigenstates $k,l=1,2$.}
\end{figure}

\begin{figure}
\centering
\includegraphics[width=.6\columnwidth]{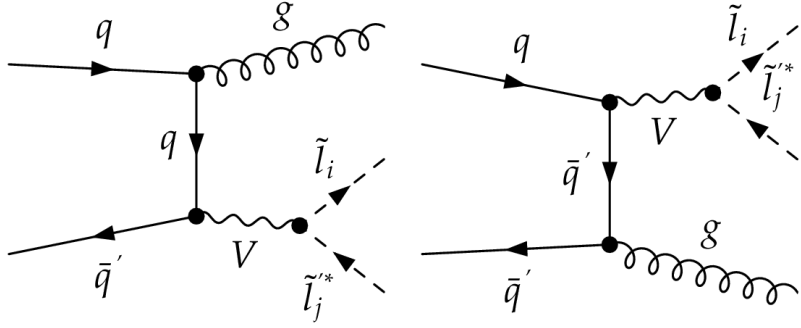}
\caption{\label{fig:3}Contributions from real gluon emission diagrams for
 slepton-pair ($V=\gamma, Z^0$) and slepton-sneutrino associated ($V=W^\mp$)
 production at next-to-leading order in perturbative QCD.}
\end{figure}

\begin{figure}
\centering
\includegraphics[width=.6\columnwidth]{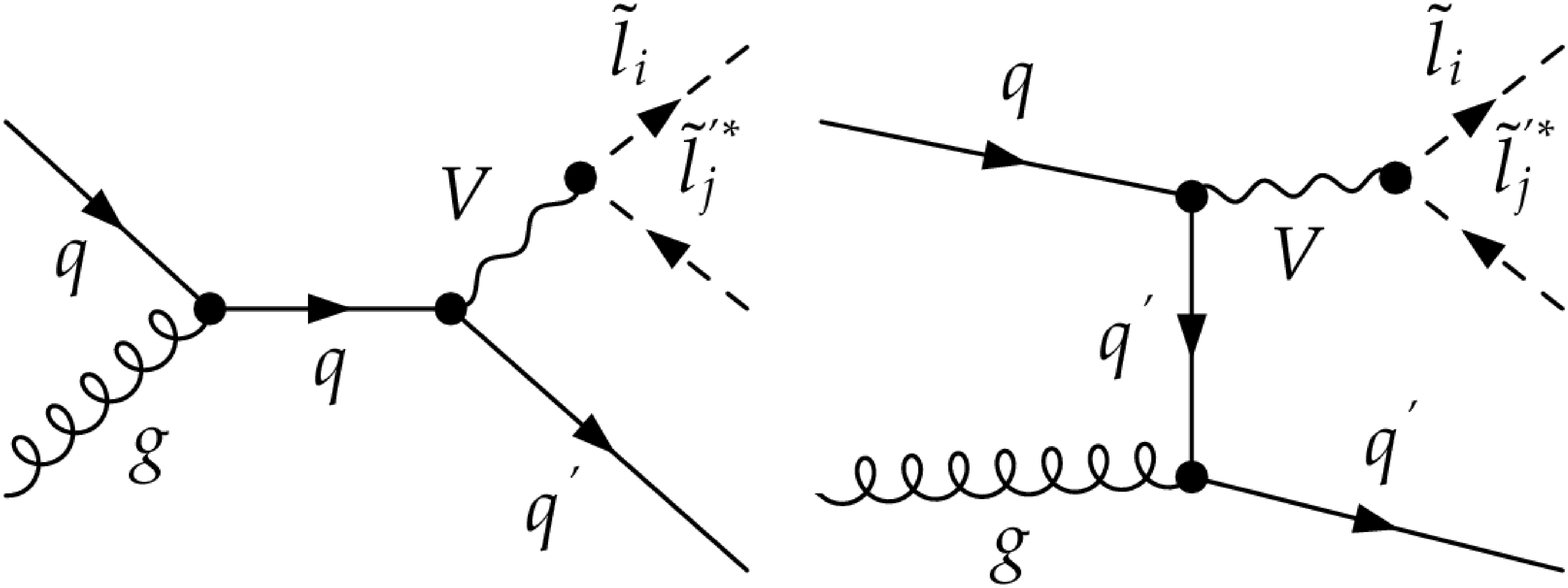}
\caption{\label{fig:4}Contributions from $qg$ diagrams for slepton-pair
 ($V=\gamma, Z^0$) and slepton-sneutrino associated ($V=W^\mp$) production
 at next-to-leading order in perturbative QCD.}
\end{figure}

The NLO QCD and SUSY-QCD corrections to the slepton-pair production cross
section have been studied for non-mixing sleptons in \cite{Baer:1997nh,%
Beenakker:1999xh}. At NLO in perturbative QCD, the quark-antiquark
annihilation process receives contributions from virtual gluon exchange (see
upper part of Fig.\ \ref{fig:2})
%
%
and real gluon emission (see Fig.\ \ref{fig:3}) diagrams, and we also have
%
%
to take into account the quark-gluon initiated subprocess (see Fig.\
\ref{fig:4}).
%
%
The infrared and collinear singularities of the three-parton cross sections
are extracted using the dipole subtraction formalism \cite{Catani:1996vz},
and the virtual corrections have been evaluated in the $\overline{{\rm MS}}$
renormalization scheme. For the SM QCD diagrams one has the well-known
results \cite{Baer:1997nh,Furmanski:1981cw}
\bea
\label{eq:qqbarnlo}
 \sigma_{q\bar{q}^{(\prime)}}^{(1;{\rm QCD})}\lr z,M^2;\frac{M^2}{\mu_F^2},
 \frac{M^2}{\mu_R^2}\rr &=&
 \sigma_0^{(\prime)}(M^2)\, C_F\, \Bigg[\left(\frac{\pi^2}{3} - 4\right)\,
 \delta (1-z) + 4 \left( \frac{\ln (1-z)}{1-z}\right)_+ - \frac{1+z^2}{1-z}
 \, \ln z \nonumber \\
 & & \hspace*{20mm}-\,2\, (1+z)\,\ln (1-z)+\frac{2\,P_{qq}^{(0)}(z)}{C_F}\,
 \ln{\frac{M^2}{\mu_F^2}} \Bigg]~~{\rm and} \\
\label{eq:qgnlo}
 \sigma_{qg}^{(1;{\rm QCD})}\lr z,M^2;\frac{M^2}{\mu_F^2},\frac{M^2}
 {\mu_R^2}\rr &=&
 \sigma_0^{(\prime)}(M^2)\, T_R\, \Bigg[\left(\frac{1}{2}-z+z^2\right)\ln
 \frac{(1-z)^2}{z} + \frac{1}{4} + \frac{3z}{2} - \frac{7 z^2}{4} +
 \frac{P_{qg}^{(0)}(z)}{T_R}\ln\frac{M^2}{\mu_F^2} \Bigg],
\eea
which expose the LO cross-sections $\sigma_0^{(\prime)}(M^2)$ in factorized
form. $C_F=4/3$ and $T_R=1/2$ are the usual QCD colour factors, and
$P_{qq,qg}^{(0)}$ are the Altarelli-Parisi splitting functions
\cite{Altarelli:1977zs}
\bea
 P_{qq}^{(0)}(z) &=& \frac{C_F}{2}\left[\frac{3}{2}\,\delta (1-z) +
 \frac{2}{(1-z)_+} - (1+z)\right]~~{\rm and} \\
 P_{qg}^{(0)}(z) &=& \frac{T_R}{2} \left[z^2 + (1-z)^2\right].
\eea
We remind the reader that our normalization corresponds to a perturbative
expansion in powers of $\alpha_s/\pi$.

The three lower diagrams of Fig.\ \ref{fig:2} contain SUSY-QCD corrections.
Generalizing results from \cite{Beenakker:1999xh} to the case of mixed
squark mass eigenstates $k,l=1,2$ in the virtual loop diagrams, we obtain
for neutral and charged currents
\bea
 \sigma_{q\bar{q}}^{(1;{\rm SUSY})}\lr z, M^2;\frac{M^2}{\mu_F^2},
 \frac{M^2}{\mu_R^2}\rr &=&
 \frac{\alpha^2\,\pi\, C_F\, \beta^3}{36\, M^2} \Bigg[f_\gamma\, e_q^2\, e_l^2\,
 \delta_{ij} + f_{\gamma Z}\, \frac{e_q\,e_l\, \delta_{ij} {\rm Re}
 \left[L_{Z \tilde{l}_i \tilde{l}_j} + R_{Z \tilde{l}_i \tilde{l}_j}
 \right]}{4\, x_W\, (1-x_W)\, (1-m_Z^2/M^2)} \nonumber\\
\label{eq:nenlo}
 && \hspace*{12mm}+\,f_Z\,  \frac{\left|L_{Z \tilde{l}_i \tilde{l}_j}
 + R_{Z \tilde{l}_i \tilde{l}_j}\right|^2}{32\, x_W^2\, (1-x_W)^2
 (1-m_Z^2/M^2)^2}\Bigg]\, \delta(1-z) ~~{\rm and}\\
\label{eq:chnlo}
 \sigma_{q\bar{q}^\prime}^{(1;{\rm SUSY})}\lr z, M^2;\frac{M^2}{\mu_F^2},
 \frac{M^2}{\mu_R^2}\rr &=&
 \frac{\alpha^2\,\pi\, C_F\, \beta^3}{36\, M^2} \Bigg[ f_W\, \frac{\left|L_{W
 \tilde{l}_i\tilde{\nu}_l} \right|^2}{32\, x_W^2\,(1-x_W)^2
 (1-m_W^2/M^2)^2}\Bigg]\, \delta(1-z),
\eea
where now only the diagonal squared photon contribution to the Born cross
section factorizes. The virtual loop coefficients $f_\gamma$, $f_{\gamma
Z}$, $f_Z$ and $f_W$ are given in App.\ \ref{sec:b}. Note that the quark
mass, which appears in the off-diagonal mass matrix elements of the squarks
running in the loops, corresponds to a linear Yukawa coupling in the
superpotential and can not be neglected, even if it is much smaller than the
total center-of-mass energy of the colliding partons allowing for a massless
factorization inside the outer hadrons.

The full NLO contributions to the cross section are then given by
\bea
\label{eq:nloqq}
 \sigma_{q\bar{q}^{(\prime)}}^{(1)}\lr z,M^2;\frac{M^2}{\mu_F^2},
 \frac{M^2}{\mu_R^2}\rr &=&
 \sigma_{q\bar{q}^{(\prime)}}^{(1;{\rm QCD})}\lr z, M^2;\frac{M^2}{\mu_F^2},
 \frac{M^2}{\mu_R^2}\rr +
 \sigma_{q\bar{q}^{(\prime)}}^{(1;{\rm SUSY})}\lr z,M^2;\frac{M^2}{\mu_F^2},
\frac{M^2}{\mu_R^2}\rr~~{\rm and}\\
\label{eq:nloqg}
 \sigma_{q g}^{(1)}\lr z, M^2;\frac{M^2}{\mu_F^2},\frac{M^2}{\mu_R^2}\rr
 &=& \sigma_{q g}^{(1;{\rm QCD})}\lr z,M^2; \frac{M^2}{\mu_F^2},\frac{M^2}
 {\mu_R^2}\rr.
\eea

\section{Threshold Resummation at Next-to-Leading Logarithmic Order}
\label{sec:3}

In this section, we recall some well-known results about soft-gluon
resummation for the Drell-Yan process. The hadronic cross section in Eq.\
(\ref{eq:cross1}) can be written in factorized form in Mellin $N$-space as
\beq
 \label{eq:mel} \sigma(N,M^{2}) = \sum_{ab} f_{a/h_a}(N+1, \mu^2_F)\,
 f_{b/h_b}(N+1, \mu^2_F)\, \hat\sigma_{ab}(N,\alpha_s, M^2/\mu^2_R,
 M^2/\mu^2_F),
\eeq
where the $N$-moments of the various quantities are defined according to the
Mellin transform
\beq
 F(N) = \int_{0}^{1} {\rm d}y\, y^{N-1}\, F(y)~,~
\eeq
with $y=\tau$, $z$, and $x_{a,b}$, respectively, for $F=\sigma$, $\hat
\sigma$, and $f_{a,b}$.

\subsection{Soft-Gluon Resummation in Mellin $N$-Space}
\label{sec:3a}

The terms leading to finite and singular contributions in the threshold
region are those proportional to $\delta(1-z)$ and to $\ln(1-z)$ and the
plus-distributions in Eqs.\ (\ref{eq:qqbarnlo}) and (\ref{eq:qgnlo}). They
respectively give rise to constant ($N$-independent) and to $(\ln N)/N$ and
$\ln^{i}N (i=1,2)$ terms in Mellin space. [For a derivation of these
correspondences we refer the reader, for instance, to App.\ A of
\cite{Catani:2003zt}.] These are the contributions that need to be resummed
to all orders in perturbative QCD. In the following, we suppress the
dependence on $M^{2}/\mu^{2}_{R}$  and $M^{2}/\mu^{2}_{F}$ for brevity.

The moments of the partonic cross section can be written in resummed form as
\cite{Sterman:1986aj, Catani:1989ne}
\beq
 \label{eq:resum}
 \hat\sigma^{({\rm RES})}_{a b}(N, \alpha_s) = \sigma_0^{(\prime)}\,
 C_{a b}(\alpha_s)\, \exp\Big[ S(N,\alpha_s)\Big].
\eeq
The $N$-independent terms are collected in the $C_{a b}$ functions
\bea
 \label{eq:coeffqq}
 C_{q\bar{q}^{(\prime)}}(\alpha_s) &=& 1 + \sum_{n=1}^{\infty}\lr
 \frac{\alpha_s}{\pi}\rr^{n}\, C^{(n)}_{q \bar{q}^{(\prime)}},\\
 \label{eq:coeffqg} C_{q g}(\alpha_{s}) &=& \sum_{n=1}^{\infty} \lr
 \frac{\alpha_{s}}{\pi} \rr^{n}\, C^{(n)}_{q g}.
\eea
These contributions are mostly due to hard virtual corrections to the cross
section, i.e.\ the terms proportional to $\delta(1-z)$ in Eqs.\
(\ref{eq:qqbarnlo}) and (\ref{eq:qgnlo}). The exponential form factor $S$
can be written at NLL as
\beq
 \label{eq:sud}
 S(N,\alpha_{s})=2\, \int_{0}^{1}{\rm d}z \frac{z^{N-1} - 1}{1 - z}
 \int_{\mu^2_F}^{(1-z)^2\,M^2} \frac{{\rm d}q^2}{q^2}\, A(\alpha_s(q^2)).
\eeq
The integrand in Eq.\ (\ref{eq:sud}) embodies the contributions coming from
the collinear emission of soft gluons from initial-state partons, i.e.\ the
terms proportional to the plus-distributions in Eq.\ (\ref{eq:qqbarnlo}). It
is a series expansion in the strong coupling constant,
\beq
 \label{eq:acoeff}
 A(\alpha_s) = \sum_{n=1}^{\infty}\lr \frac{\alpha_{s}}{\pi}\rr^{n}\,
 A_{n},
\eeq
whose coefficients are perturbatively computable through a fixed-order
calculation. In particular, it has been proven \cite{Korchemsky:1988si} that
in the $\overline{\mathrm {MS}}$ factorization scheme the coefficients of
the $A$-function are exactly equal to the large-$N$ coefficients of the
diagonal splitting function
\beq
 \gamma_{qq}(\alpha_s)~=~
 \int_{0}^{1} {\rm d}z\, z^{N-1}\, P_{qq}(z)~=~
 -A(\alpha_s)\, \ln\bar{N} + \mathcal{O}(1),
\eeq
where $\bar{N} = N\,\exp[\gamma_E]$ and $\gamma_{E}$=0.5772... is the
Euler-Mascheroni constant. Performing the integration in Eq.\ (\ref{eq:sud})
and using Eq.\ (\ref{eq:acoeff}), we obtain the form factor up to NLL,
\beq
 S(N, \alpha_s) = g_1(\lambda)\, \ln \bar{N} + g_2(\lambda).
\eeq
The functions $g_1$ and $g_2$ resum the LL ($\alpha_s^n\ln^{n+1} N$) and NLL
($\alpha_s^n \ln^n N$) contributions, respectively, and are given by
\cite{Sterman:1986aj, Catani:1989ne}
\bea
 g_1(\lambda) &=& \frac{A_1}{\beta_0\lambda}\, \le 2\, \lambda + (1 - 2\,
 \lambda)\ln(1 - 2\, \lambda) \re~~~{\rm and}\\
 g_2(\lambda) &=& \frac{A_1\beta_1}{\beta_0^3}\, \le 2\, \lambda +
 \ln(1 - 2\, \lambda) + \frac{1}{2}\ln^2(1 - 2\,\lambda) \re -
 \frac{A_2}{\beta_0^2}\le 2\, \lambda + \ln(1 - 2\, \lambda)\re\nonumber\\
 &+& \frac{A_1}{\beta_0}\le 2\, \lambda + \ln(1 - 2\, \lambda)\re
 \ln\frac{M^2}{\mu^2_R} - \frac{2\, A_{1}\, \lambda}{\beta_0} \ln\frac{M^2}
 {\mu^2_F},
\eea
where $\lambda = [\beta_0\, \alpha_s\, \ln \bar{N}]/\pi$. The first two
coefficients of the QCD $\beta$-function are
\beq
 \beta_0 = \frac{1}{12}(11\, C_A - 2\, N_f) ~~{\rm and}~~ \beta_1 =
 \frac{1}{24}(17\, C^2_A - 5\, C_A\, N_f - 3\, C_F\, N_f),
\eeq
$N_f$ being the number of effectively massless quark flavours and $C_F =
4/3$, $C_A = 3$ the usual QCD colour factors.

Thus, the knowledge of the first two coefficients of the function
$A(\alpha_s)$ \cite{Kodaira:1981nh, Catani:1988vd},
\beq
 A_1 = C_F ~~{\rm and}~~ A_2 = \frac{1}{2}\, C_F\, \le C_A\lr\frac{67}{18} -
 \frac{\pi^2}{6}\rr - \frac{5}{9}\, N_f\re,
\eeq
together with the first coefficients of the $C$-functions in Eqs.\
(\ref{eq:coeffqq}) and (\ref{eq:coeffqg}),
\bea
 \label{eq:c1} C^{(1)}_{q\bar{q}^{(\prime)}} &=& C_F\lr\frac{2\, \pi^2}{3}
 - 4 + \frac{3}{2}\ln\frac{M^2}{\mu^2_F}\rr ~~{\rm and}~~ \\
 \label{eq:c2} C^{(1)}_{q g} &=& 0,
\eea
allows us to perform resummation up to NLL.

\subsection{Improvements of the Resummation Formalism}
\label{sec:3b}

In the limit of large $N$, the cross section is clearly dominated
by terms of $\mathcal{O}(\ln^{2} N)$, $\mathcal{O}(\ln N)$ and
$\mathcal{O}(1)$. It seems thus reasonable to neglect terms
suppressed by powers of $1/N$ in the resummation formalism.
Actually these last terms are multiplied by powers of $\ln N$ and
could as well provide a non-negligible effect in the threshold
limit. In \cite{Kramer:1996iq, Catani:2001ic} it has been shown
that these contributions are due to collinear parton emission and
can be consistently included in the resummation formula, leading
to a ``collinear-improved'' resummation formalism. The
modification simply amounts to the introduction of an
$N$-dependent term in the $C^{(1)}_{q \bar{q}^{(\prime)}}$ and
$C^{(1)}_{q g}$ coefficient of Eqs.\ (\ref{eq:c1}) and
(\ref{eq:c2})
\bea
 \label{eq:c1p}
 C^{(1)}_{q \bar{q}^{(\prime)}}&\rightarrow&
 \widetilde{C}^{(1)}_{q \bar{q}^{(\prime)}} =
 C^{(1)}_{q \bar{q}^{(\prime)}} + 2\, A_{1}\, \frac{\ln{\bar{N}} -
 \frac{1}{2} \ln{\frac{M^2}{\mu_f^2}}}{N}~, \\
 \label {eq:c2p}
 C^{(1)}_{q g} &\rightarrow& \widetilde{C}^{(1)}_{q g} = C^{(1)}_{qg} -
 T_R\,\frac{\ln{\bar{N}} - \frac{1}{2}\ln{\frac{M^2}{\mu_f^2}}}{N}.
\eea

Furthermore, the exponentiation of the contributions embodied in
the $C$-function has been proved in \cite{Eynck:2003fn}, leading
to the following modification in Eq.\ (\ref{eq:resum}):
\beq
 \label{eq:impr}
 \hat\sigma^{({\rm RES})}_{a b}(N,\alpha_{s}) = \sigma_0^{(\prime)}\,
 \exp\Big[C^{(1)}_{q\bar{q}^{(\prime)}}(\alpha_s)\Big]\,\exp
 \Big[S(N,\alpha_{s})\Big]~.
\eeq
As the authors of Ref.\ \cite{Eynck:2003fn} recognize, this exponentiation
of the $N$-independent terms is not comparable to the standard threshold
resummation in terms of predictive power. While in the latter case
a low-order calculation can be used to predict the behaviour of
full towers of logarithms, in the former case it is not possible
to directly get information on the behaviour of constant terms at,
say, $n$ loops, but a complete calculation at the $n^{{\rm th}}$
perturbative order will still be necessary. Nonetheless, the
comparison of the numerical results obtained with and without the
exponentiation of the constant terms can at least provide an
estimate of the errors due to missing higher-order corrections.

\subsection{Inverse Mellin-Transform and Matching Procedure}
\label{sec:3c}

Once resummation has been achieved in $N$-space, an inverse Mellin-transform
back to the physical $x$-space is needed. The customary way to perform this
inversion, avoiding the singularities of the $N$-moments, is the ``Minimal
Prescription'' of \cite{Catani:1996yz},
\beq
 \sigma=\frac{1}{2\, \pi\, i}\int_{C_{MP} - i\, \infty}^{C_{MP} +
 i\, \infty} {\rm d}N \lr\frac{M^2}{S} \rr^{-N} \sigma(N, M^2).
\eeq
The constant $C_{MP}$ has to be chosen so that all the poles in the
integrand are to the left of the integration contour in the complex
$N$-plane except for the Landau pole at $N = \exp[\pi/(2\, \beta_0\,
\alpha_s)]$, which should lie far to the right on the real axis.

Finally, a matching procedure of the NLL resummed cross section to the NLO
result has to be performed in order to keep the full information contained
in the fixed-order calculation and to avoid possible double-counting of the
logarithmic enhanced contributions. A correct matching is achieved through
\beq
 \label{eq:total}
 \sigma = \sigma^{({\rm F.O.})} + \frac{1}{2\,  \pi\, i} \int_{C_{MP} -
 i\, \infty}^{C_{MP} + i\, \infty} {\rm d}N \lr\frac{M^2}{S} \rr^{-N} \le
 \sigma^{({\rm RES})}(N, M^2) - \sigma^{({\rm EXP})}(N, M^2)\re,
\eeq
where $\sigma^{({\rm F.O.})}$ is the fixed-order perturbative result, $\sigma^
{({\rm RES})}$ is the resummed cross section, and $\sigma^{({\rm EXP})}$ is the
truncation of the resummed cross section to the same perturbative order as
$\sigma^{({\rm F.O.})}$. In our case the expansion of the resummed
partonic cross section up to order $\alpha_s$ reads
\bea
 \label{eq:expqq}
 \hat\sigma^{({\rm EXP})}_{q \bar{q}^{(\prime)}}(N,M^{2}) &=& \sigma_0^{(\prime)}
 (M^2) \le 1 + \frac{\alpha_{s}}{\pi}\, \lr C_{F}\,\Big( 2\, \ln^2\bar{N} -
 2\, \ln \bar{N} \ln\frac{M^{2}}{\mu^{2}_{F}} \Big) +
 \widetilde{C}^{(1)}_{q \bar{q}^{(\prime)}} \rr\re + \mathcal{O}
 (\alpha^{2}_{s})~~{\rm and} \\
 \label{eq:expqg}
 \hat\sigma^{({\rm EXP})}_{q g}(N, M^2) &=&\sigma_0^{(\prime)}(M^2) \le
 \frac{\alpha_{s}}{\pi}\,\widetilde{C}^{(1)}_{q g} \re + \mathcal{O}
 (\alpha^{2}_{s}).
\eea
In Mellin-space, the fixed order NLO cross sections of Eqs.\
(\ref{eq:qqbarnlo}) and (\ref{eq:qgnlo}) read \cite{Martin:1997rz}
\bea
 \hat\sigma^{({\rm F.O.})}_{q \bar{q}^{(\prime)}}(N, M^2)&=&
 \sigma_0^{(\prime)}(M^2) \Bigg[ 1 + \frac{\alpha_{s}}{\pi}\, C_{F}\,
 \left( 4\, S_1^2(N) - \frac{4}{N\,(N+1)}\, S_1(N) + \frac{2}{N^2} +
 \frac{2}{(N+1)^2} \right. \nonumber \\
 &-& \left. 8 + \frac{4\,\pi^2}{3} + \left[ \frac{2}{N\, (N+1)} + 3 - 4\,
 S_1(N)\right] \,\ln \frac{M^2}{\mu_F^2} \right)\Bigg] +
 \mathcal{O}(\alpha^{2}_{s})~~{\rm and}\\
 \label{eq:exaqg}
 \hat\sigma^{({\rm F.O.})}_{q g}(N, M^2) &=& \sigma_0^{(\prime)}(M^2)
 \Bigg[ \frac{\alpha_{s}}{\pi}\,T_R\, \left( -2\, \frac{N^2 + N +
 2}{N\, (N+1) (N+2)}\, S_1(N) + \frac{N^4 + 11\, N^3 + 22\, N^2 +
 14\,N + 4}{N^2\, (N+1)^2 (N+2)^2} \nonumber \right.\\
 &+&\left. \frac{N^2+N+2}{N\, (N+1) (N+2)}\, \ln\frac{M^2}{\mu_F^2} \right)
 \Bigg] + \mathcal{O}(\alpha^{2}_{s})
\eea
with $S_1(N)=\sum_{j=1}^N 1/j$. In the large-$N$ limit, $S_1(N) \simeq
\ln\bar{N} + 1/(2\,N)$, and we get
\bea
 \label{eq:exaqq}
 \hat\sigma^{({\rm F.O.})}_{q \bar{q}^{(\prime)}}(N,M^{2}) &=& \sigma_0^{(\prime)}
 (M^2) \le 1 + \frac{\alpha_{s}}{\pi}\, \lr C_{F}\,\Big( 2\, \ln^2\bar{N} -
 2\, \ln \bar{N} \ln\frac{M^{2}}{\mu^{2}_{F}} \Big) + \widetilde{C}^{(1)}_{q
 \bar{q}^{(\prime)}} \rr\re + \mathcal{O}(\alpha^{2}_{s})~~{\rm and} \\
 \label{eq:exaqg}
 \hat\sigma^{({\rm F.O.})}_{q g}(N, M^2) &=& \sigma_0^{(\prime)}(M^2) \le
 \frac{\alpha_{s}}{\pi}\, \widetilde{C}^{(1)}_{q g} \re + \mathcal{O}
 (\alpha^{2}_{s}).
\eea
Comparing Eqs.\ (\ref{eq:expqq}), (\ref{eq:expqg}), (\ref{eq:exaqq}), and
(\ref{eq:exaqg}), we see that the expansion of the resummed cross section at
order $\alpha_s$ correctly reproduces the fixed order result in the
large-$N$ limit, including even terms that are suppressed by $1/N$.

\section{Numerical Results}
\label{sec:4}

For the masses and widths of the electroweak gauge bosons, we use the
current values of $m_Z=91.1876$ GeV, $m_W=80.403$ GeV, $\Gamma_Z=2.4952$
GeV, and $\Gamma_W=2.141$ GeV. The CKM-matrix elements are computed using
\bea
 V = \left(\begin{array}{c c c}
  c_{12} c_{13} & s_{12} c_{13} & s_{13} e^{-i \delta}\\
 -s_{12} c_{23} - c_{12} s_{23} s_{13} e^{i \delta}& c_{12} c_{23} - s_{12}
 s_{23} s_{13} e^{i \delta}& s_{23} c_{13}\\
 s_{12} s_{23} - c_{12} c_{23} s_{13} e^{i \delta}& -c_{12} s_{23} - s_{12}
 c_{23} s_{13} e^{i \delta}& c_{23} c_{13}
 \end{array}\right)
\eea
with $s_{ij}=\sin\theta_{ij}$ and $c_{ij}=\cos\theta_{ij}$, $\theta_{ij}$
being the usual angles relative to the mixing of two specific generations
$i$ and $j$ and $\delta$ being the CP-violating complex phase. Their average
values are given by
\bea
 s_{12} = 0.2243~,~s_{23} = 0.0413 ~,~ s_{13} = 0.0037 ~,~~{\rm and}~~
 \delta = 1.05.
\eea
The squared sine of the electroweak mixing angle
\bea
 \sin^2\theta_W &=& 1-m_W^2/m_Z^2
\eea
and the electromagnetic fine structure constant
\bea
 \alpha &=& \sqrt{2} G_F m_W^2 \sin^2\theta_W / \pi
\eea
can be calculated in the improved Born approximation using the world average
value of $G_F=1.16637\cdot 10^{-5}$ GeV$^{-2}$ for Fermi's coupling
constant \cite{Yao:2006px}.

The physical masses of the SUSY particles and the mixing angles are computed
with the computer program SUSPECT \cite{Djouadi:2002ze}, including a
consistent calculation of the Higgs mass, with all one-loop and the dominant
two-loop radiative corrections in the renormalization group equations that
link the restricted set of SUSY-breaking parameters at the gauge coupling
unification scale to the complete set of observable SUSY masses and mixing
angles at the electroweak scale. We choose one minimal supergravity (mSUGRA)
point, SPS 1a, and one gauge-mediated supersymmetry breaking (GMSB) point,
SPS 7, as benchmarks for our numerical study \cite{Allanach:2002nj}. SPS 1a
is a typical mSUGRA point with an intermediate value of $\tan\beta= 10$ and
$\mu>0$. It has a model line attached to it, which is specified by
$m_0=-A_0=0.4~m_{1/2}$. For $m_{1/2}=250$ GeV, this SUSY-breaking scenario
leads to light sleptons $\tilde{\tau}_1$, $\tilde{e}_1$, $\tilde{\tau}_2$,
$\tilde{e}_2$, $\tilde{\nu}_\tau$ and $\tilde{\nu}_e$ with masses of 136.2,
146.4, 216.3, 212.3, 196.1 and 197.1 GeV and to heavy squarks with masses
around 500-600 GeV \footnote{While the top-squark mass eigenstate
$\tilde{t}_1$ is slightly lighter, it does nonetheless not contribute to
the virtual squark loops due to the negligible top-quark density in the
proton.}. SPS 7 is a GMSB scenario with a $\tilde{\tau}_1$ as the
next-to-lightest SUSY particle (NLSP) and an effective SUSY-breaking scale
$\Lambda=40$ TeV, $N_{\rm mes}=3$ messenger fields of mass $M_{\rm mes}=80$
TeV, $\tan\beta=15$, and $\mu>0$, which leads again to light sleptons with
masses of 114.8, 121.1, 263.9, 262.1, 249.5 and 249.9 GeV, respectively, and
even heavier squarks with masses around 800-900 GeV. Its model line is
defined by $M_{\rm mes} = 2\, \Lambda$. The slepton masses and mixing angles
are actually quite similar for the SPS 1a mSUGRA and SPS 7
GMSB points, so that the corresponding production cross sections will not
differ significantly. Slepton detection will, however, be slightly different
in both scenarios, as the sleptons decay to a relatively massive neutralino
($\tilde{\chi}^0 _1$) lightest SUSY particle (LSP) at SPS 1a, but to a very
light gravitino LSP at SPS 7. The lightest tau slepton thus decays into a
tau lepton and missing transverse energy. Feasibility studies of tau-slepton
identification at the LHC with the ATLAS detector \cite{Hinchliffe:2002se}
and tau tagging with the CMS detector \cite{Gennai:2002qq} have recently
shown that stau masses should be observable up to the TeV range.

Our cross sections are calculated for the Tevatron $p\bar{p}$-collider,
currently operating at $\sqrt{S}=1.96$ TeV, as well as for the LHC
$pp$-collider, bound to operate at $\sqrt{S}=14$ TeV starting in 2008. For
the LO (NLO and NLL) predictions, we use the LO 2001 \cite{Martin:2002dr}
(NLO 2004 \cite{Martin:2004ir}) MRST-sets of parton distribution functions
(PDFs). For the NLO and NLL predictions, $\alpha_s$ is evaluated with the
corresponding value of $\Lambda_{\overline{\rm MS}}^{n_f=5}=255$ MeV at
two-loop accuracy. We fix the unphysical scales $\mu_{F}$ and $\mu_{R}$
equal to the invariant mass $M$ of the slepton (slepton-sneutrino) pair.

\subsection{Invariant-Mass Distributions for Slepton Pairs}
\label{sec:4a}

%
\begin{figure}
\centering
\includegraphics[width=.7\columnwidth]{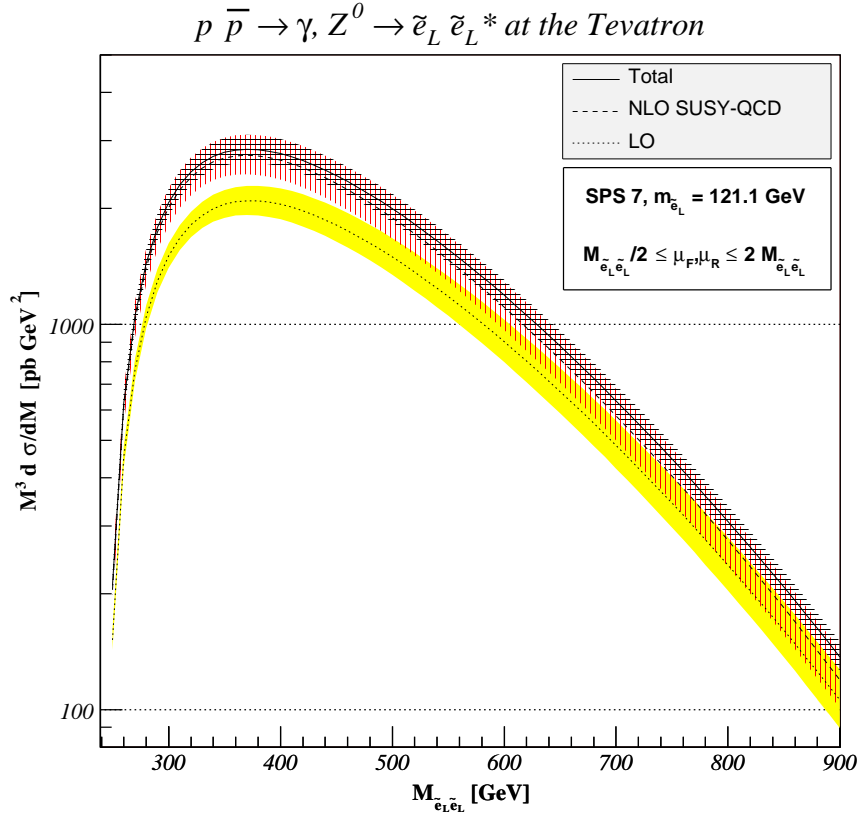}
\includegraphics[width=.7\columnwidth]{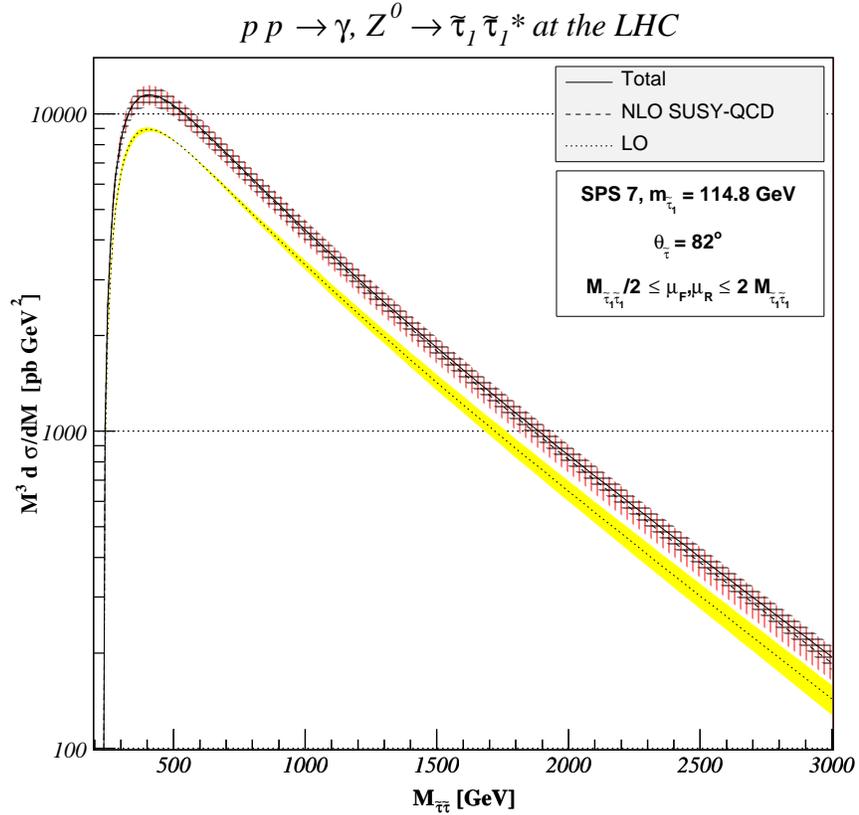}
\caption{\label{fig:5}Invariant-mass distribution $M^3\,\d\sigma/\d M$
 of $\tilde{e}_L$-pairs at the Tevatron (top) and $\tilde{\tau}_1$-pairs
 at the LHC (bottom) for the benchmark point SPS 7. We show the total
 NLL+NLO matched and the fixed order NLO SUSY-QCD and LO QCD results,
 including the respective scale uncertainties as horizontally hatched,
 vertically hatched and shaded bands.}
\end{figure}
%
The invariant-mass distribution $M^3\d\sigma/\d M$ for first- (and
equal-mass second-) generation sleptons at the Tevatron is shown in Fig.\
\ref{fig:5} (top), the one for (slightly lighter) third-generation sleptons
at the LHC in Fig.\ \ref{fig:5} (bottom). In both cases, we have chosen the
SPS 7 GMSB benchmark point. The differential cross section $\d\sigma/\d M$
has been multiplied by a factor $M^3$ in order to remove the leading mass
dependence of propagator and phase space factors. As is to be expected for
$P$-wave production of scalar particles, the distributions rise above the
threshold at $\sqrt{S}=2 m_{\tilde{l}}$ with the third power of the slepton
velocity $\beta$, see Eq.\ (\ref{eq:sig0n}), and peak at about 100 GeV above
threshold (at 370 GeV for $M^3\d\sigma/\d M$ and 310 GeV for $\d\sigma/\d
M$ for the Tevatron; 410 GeV and 300 GeV for the LHC), before falling off
steeply due to the $s$-channel propagator and the decreasing parton
luminosity. As can also be seen from Eqs.\ (\ref{eq:qqbarnlo}) and
(\ref{eq:qgnlo}), the QCD corrections do not alter the $P$-wave velocity
dependence close to threshold.
At the Tevatron, the total and NLO SUSY-QCD predictions exceed the maximal
LO cross section by 36 and 31\%, respectively, whereas at the LHC, the
maximal cross section increases by 28 and 27\%. Threshold resummation
effects are thus clearly more important at the Tevatron, where the hadronic
center-of-mass energy is limited and the scaling variable $\tau=M^2/S$ is
closer to one, and they increase with $M$ to the right of both plots.
The maximal theoretical error is estimated in Fig.\ \ref{fig:5} by an
independent
variation of the factorization and renormalization scales between $M/2$ and
$2M$. It is indicated as a shaded, vertically, and horizontally hatched band
for the LO, NLO SUSY-QCD, and the total prediction. At LO, the only
dependence comes from the factorization scale. It increases with the
momentum-fraction $x$ of the partons in the proton or anti-proton and is
therefore already substantial for small $M$ at the Tevatron, but only for
larger $M$ at the LHC. At NLO, this dependence is reduced due to the
factorization of initial-state singularities, but a strong additional
dependence is introduced by the renormalization scale in the coupling
$\alpha_s(\mu_R)$. After resummation, this dependence is reduced as well, so
that the total scale uncertainty at the Tevatron diminishes from 20\%--35\%
for NLO to only 16\%--17\% for the matched resummed result. The reduction
is, of course, more important in the large-$M$ region. At the LHC, where
$\alpha_s$ is evaluated at a larger renormalization scale and is thus less
sensitive to it, the corresponding numbers are 18\%--25\% and 15\%--17\%.

%
\begin{figure}
\centering
\includegraphics[width=.7\columnwidth]{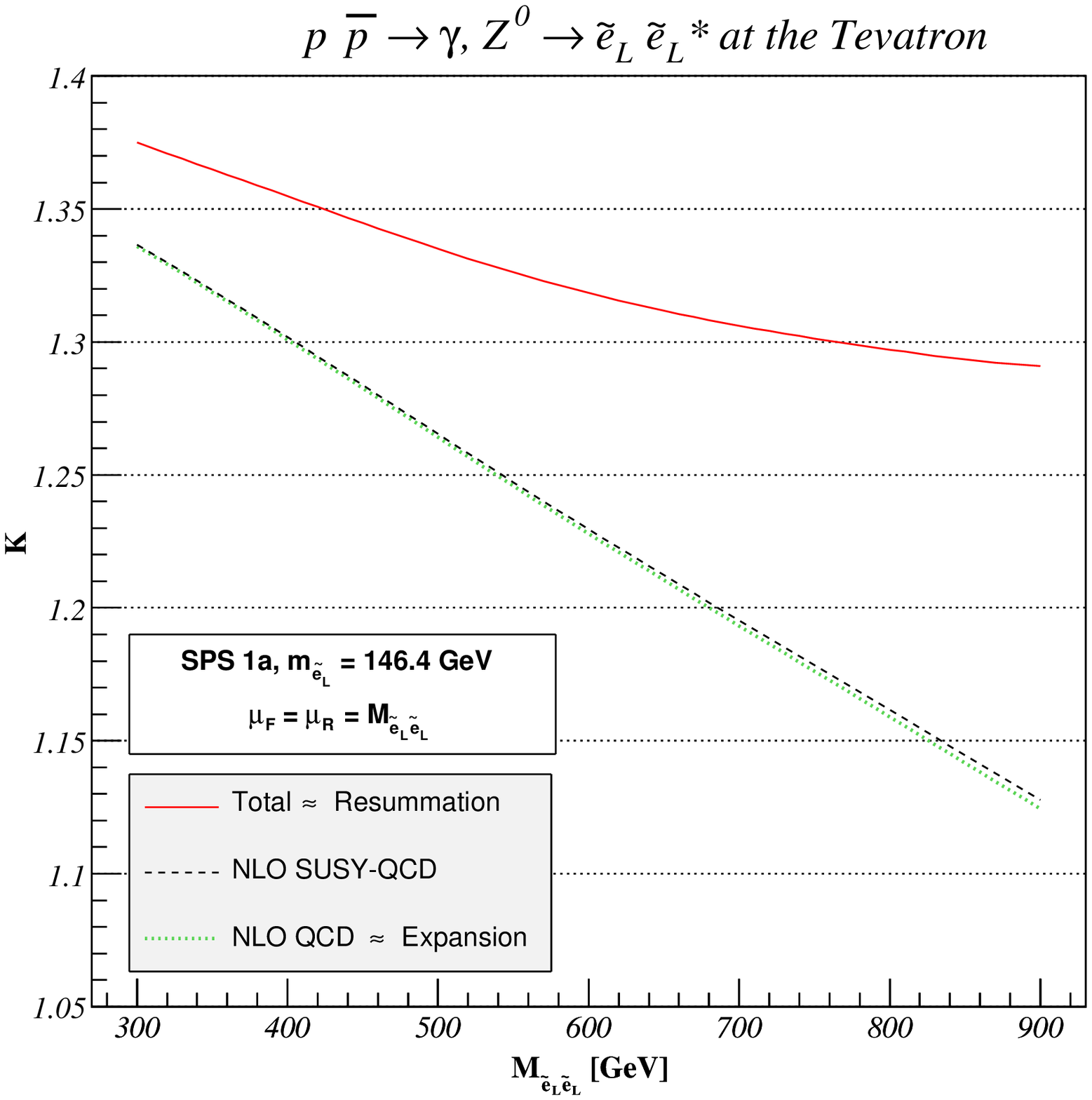}
\includegraphics[width=.7\columnwidth]{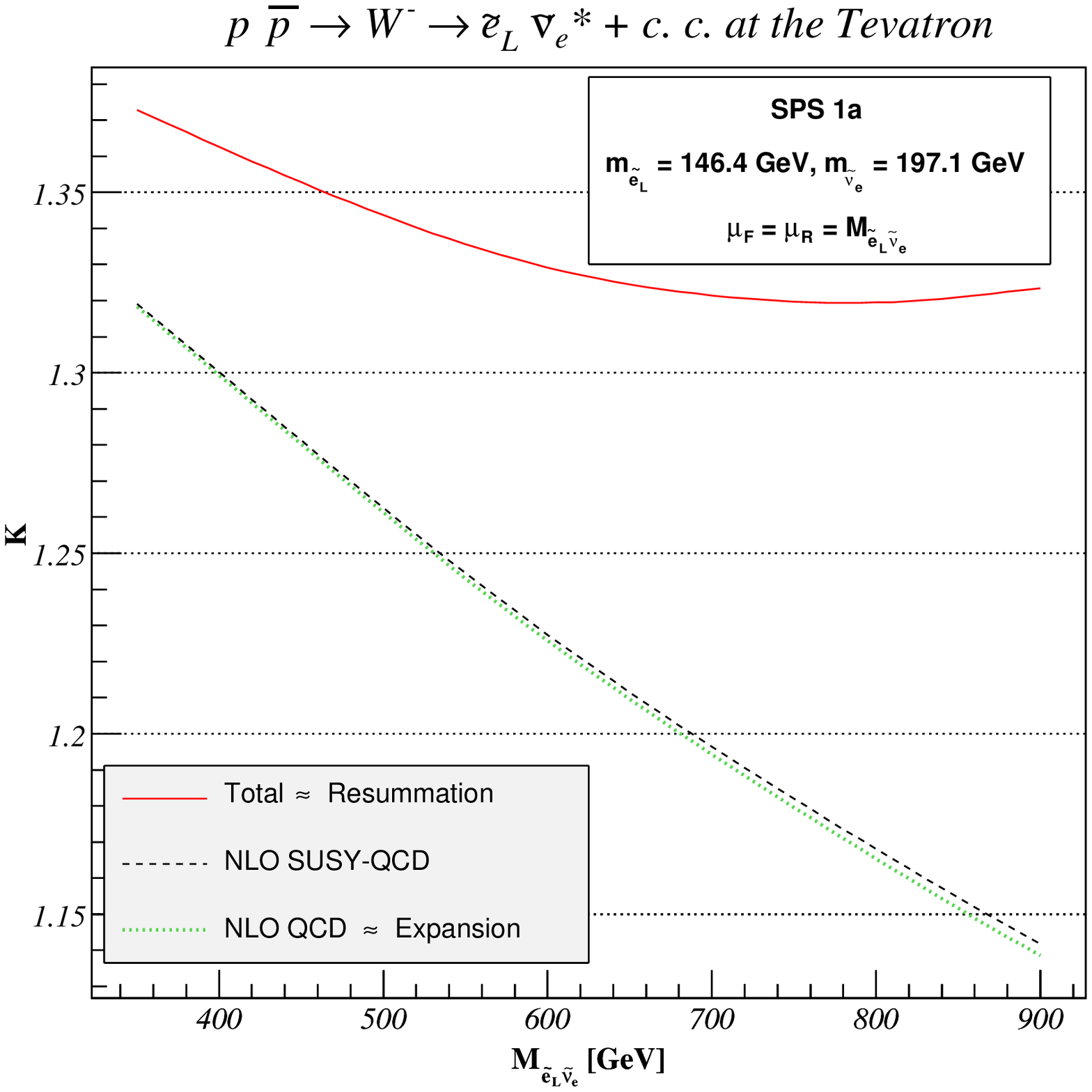}
\caption{\label{fig:6}$K$-factors as defined in Eq.\ (\ref{eq:K}) for
 $\tilde{e}_L$-pair (top) and associated $\tilde{e}_L\tilde{\nu}_e^*$
 production (bottom) at the Tevatron for the benchmark point SPS 1a. We show
 the total NLL+NLO matched result, which is almost identical to the purely
 resummed result at NLL, as well as the fixed order NLO SUSY-QCD and QCD
 results. The latter practically coincides with the resummed result expanded
 up to NLO.}
\end{figure}
%
%
\begin{figure}
\centering
\includegraphics[width=.7\columnwidth]{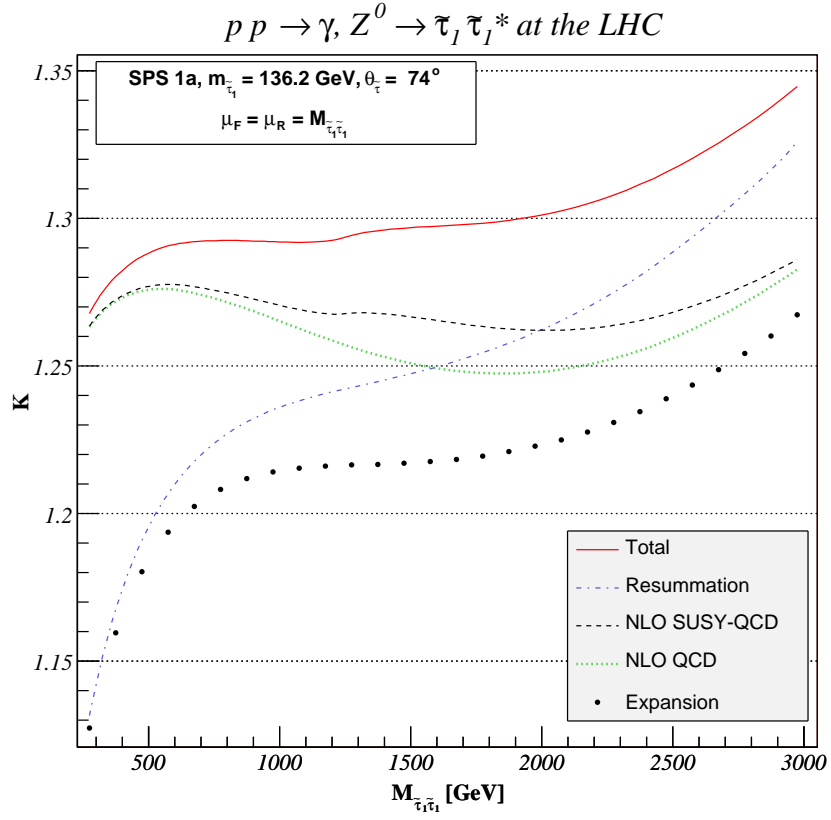}
\includegraphics[width=.7\columnwidth]{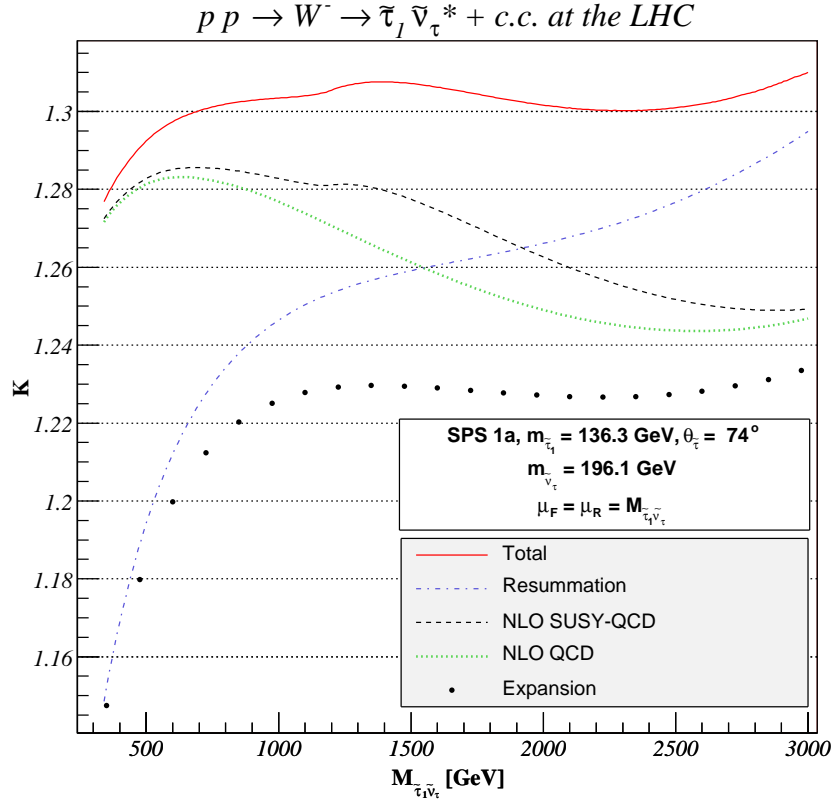}
\caption{\label{fig:7}$K$-factors as defined in Eq.\ (\ref{eq:K}) for
 $\tilde{\tau}_1$-pair (top) and associated $\tilde{\tau}_1\tilde{\nu}_\tau
 ^*$ production (bottom) at the LHC for the benchmark point SPS 1a. We show
 the total NLL+NLO matched result, the resummed result at NLL, the fixed
 order NLO SUSY-QCD and QCD results, and the resummed result expanded up to
 NLO.}
\end{figure}
%
For the mSUGRA scenario SPS 1a, with nonetheless similar slepton masses and
mixing angles (see above), we show in Figs.\ \ref{fig:6} and \ref{fig:7} the
cross section correction factors
\bea
 \label{eq:K}
 K^i = \frac{{\rm d}\sigma^i / {\rm d}M}{{\rm d}\sigma^{\rm LO} / {\rm d}M},
\eea
where $i$ labels the corrections induced by NLO QCD (Eqs.\
(\ref{eq:qqbarnlo}) and (\ref{eq:qgnlo})), additional NLO SUSY-QCD (Eqs.\
(\ref{eq:nloqq}) and (\ref{eq:nloqg})), resummation (Eqs.\ (\ref{eq:resum}),
(\ref{eq:c1p}), and (\ref{eq:c2p})), and the matched total contributions
(Eq.\ (\ref{eq:total})) as well as the fixed-order expansion (Eqs.\
(\ref{eq:expqq}) and (\ref{eq:expqg})) of the resummation contribution as a
function of the invariant mass $M$. As one can see immediately, the
mass-dependence of these corrections for charged-current associated
production of sleptons and sneutrinos (lower parts of Figs.\ \ref{fig:6} and
Fig.\ \ref{fig:7}) does not differ substantially from the mass-dependence of
the neutral-current production of slepton-pairs (upper parts).

At the Tevatron (Fig.\ \ref{fig:6}), where we are close to the threshold,
resummation effects are already important at low $M$ (4\%) and increase to
sizeable 16\% at large $M$. The NLO QCD result is thus dominated by large
logarithms and coincides with the expanded result at the permille level. In
addition, the relative importance of the (finite) SUSY-QCD contributions is
reduced, and the total prediction coincides with the resummed prediction,
since fixed-order and expanded contributions cancel each other in Eq.\
(\ref{eq:total}). We have also verified that exponentiating the finite
($N$-independent) terms collected in the coefficient function $C^{(1)}_
{q\bar{q}^{(\prime)}}$, as proposed in \cite{Eynck:2003fn} (see Eq.\
\ref{eq:impr}), leads only to a 0.6\%--0.8\% increase of the matched
resummed result. The Tevatron being a $p\bar{p}$-collider, the total cross
section is dominated by $q\bar{q}$-annihilation, and $qg$-scattering
contributes at most 1\% at small $M$ (or small $x$), where the gluon density
is still appreciable. Integration over $M$ leads to total cross sections
for the neutral (charged) current processes in Fig.\ \ref{fig:6} of 4.12
(3.92) fb in LO, 5.3 (4.96) fb in NLO (SUSY-)QCD, and 5.55 (5.28) fb for the
matched resummed calculation. The corresponding (global) $K$-factors
\bea
 \label{eq:kglob} K^i_{{\rm glob}} &=& \frac{\sigma^{i~~}} {\sigma^{\rm LO}}
 ~=~ \frac{\int {\rm d}M\, {\rm d}\sigma^{i~~} / {\rm d}M}
 {\int {\rm d}M\,{\rm d}\sigma^{\rm LO} / {\rm d}M}
\eea
are then 1.29 (1.27) at fixed-order and 1.35 (1.35) with resummation.

At the LHC (Fig.\ \ref{fig:7}), sleptons can be produced with relatively
small invariant mass $M$ compared to the total available center-of-mass
energy $\sqrt{S}$, so that $z=\tau/(x_ax_b)=M^2/s\ll 1$ and the resummation
of ($1-z$)-logarithms is less important. This is particularly true for the
production of the light mass-eigenstates of mixing third-generation
sleptons, as shown in Fig.\ \ref{fig:7}. In the low-$M$ (left) parts of
these plots, the total result is less than 0.5\% larger than the NLO
(SUSY-)QCD result. Only at large $M$ the logarithms become important and
lead to a 7\% increase of the $K$-factor with resummation over the
fixed-order result. In this region, the resummed result approaches the total
prediction, since the NLO QCD calculation is dominated by large logarithms
and approaches the expanded resummed result. However, we are still far from
the hadronic threshold region, so that both resummed and fixed-order
contributions and a consistent matching of the two are needed. At low
$M$, where finite terms dominate, the resummed contribution is close to its
fixed-order expansion and disappears with $M$. In the intermediate-$M$
region, one can observe the effect of SUSY-QCD contributions, in particular
the one coming from the $\tilde{q}\tilde{q}\tilde{g}$-vertex correction
(lower left diagram in Fig.\ \ref{fig:2}). As $M \geq 2 m_{\tilde{q}}$, one
crosses the threshold for squark-pair production and observes a resonance in
Fig.\ \ref{fig:7}. As for the Tevatron, exponentiating the finite
($N$-independent) terms collected in the coefficient function $C^{(1)}_
{q\bar{q}^{(\prime)}}$ leads only to a 1\% increase of the matched
resummed result. The LHC being a high-energy $pp$-collider, it has a
significant gluon-luminosity, in particular at small $M$ (or $x$), and
indeed the $qg$-subprocess changes (lowers) the total cross section by
7\% at small $M$ and 3\% at large $M$. After integration over $M$, we obtain
total cross sections of 27 (9.59) fb in LO, 34.3 (12.3) fb in NLO SUSY-QCD,
and 34.6 (12.5) fb for the resummed-improved result, corresponding to global
$K$-factors of 1.28 for fixed-order and 1.29 for the matched resummed cross
section for both processes. Resummation of large logarithms is thus not as
important as for the Tevatron at the benchmark point SPS 1a.

\subsection{Scale Variations of the Total Cross Section}
\label{sec:4b}

In this section, we study the dependence of total, i.e. invariant-mass
integrated, slepton-pair and slepton-sneutrino cross sections on three
different scales: first the dependence on the effective SUSY-breaking scale
$\Lambda$ as defined in the GMSB model line SPS 7, and then the dependence
on the renormalization and factorization scales $\mu_{R,F}$ at the point
SPS 7. We remind the reader that the model line for SPS 7 is $M_{\rm mes} =
2\,\Lambda$ with $N_{\rm mes}=3$, $\tan\beta=15$, and $\mu>0$ fixed. The
benchmark point at $\Lambda=40$ TeV will be indicated in the figures where
$\Lambda$ varies as a vertical dashed line, and we show in addition to the
scale $\Lambda$ the mass scale of the produced charged slepton, ranging from
80 (87.5) to 280 (385) GeV for $\tilde{e}_L$ ($\tilde{\tau}_1$).

%
\begin{figure}
\centering
\includegraphics[width=.7\columnwidth]{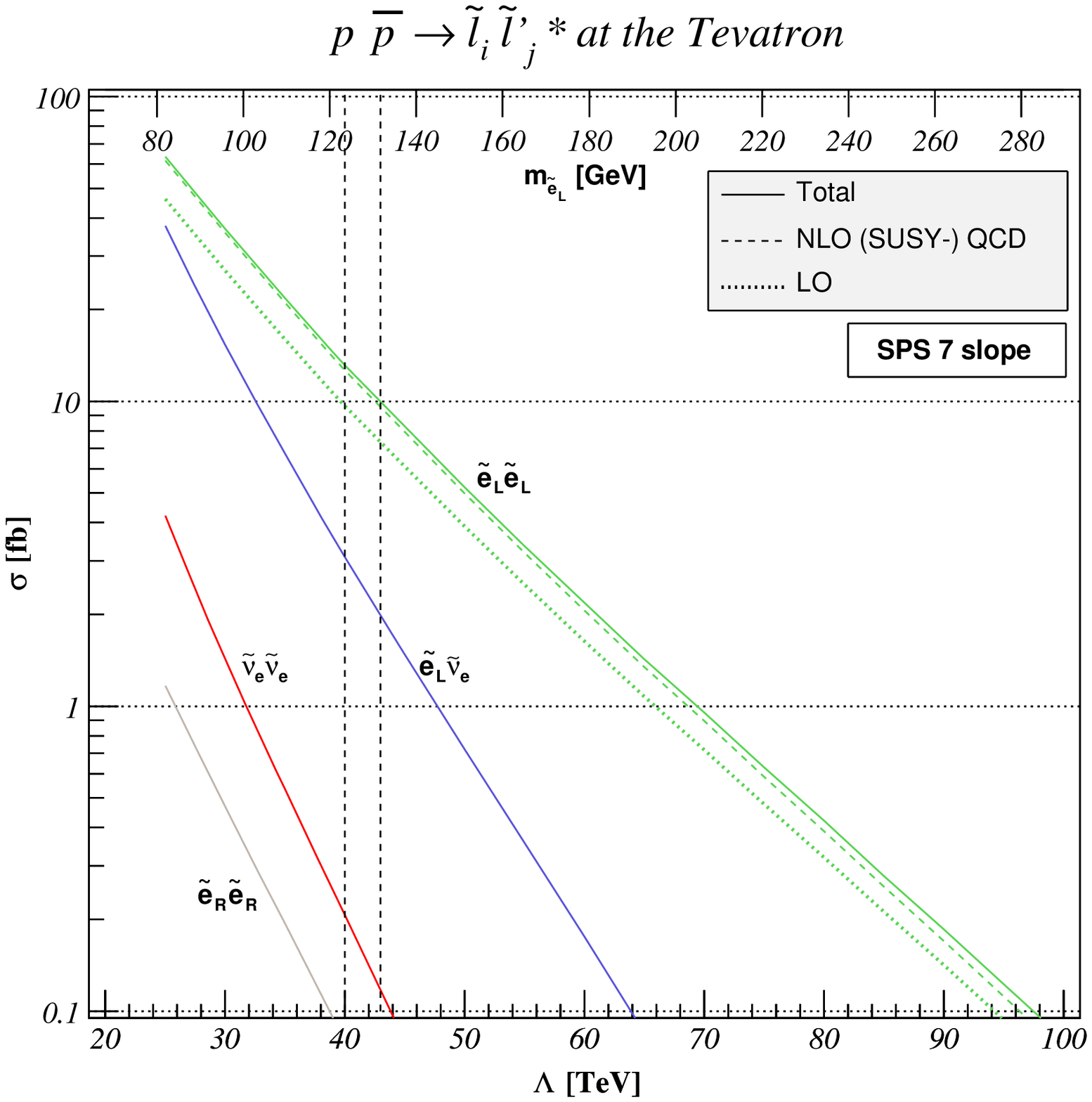}
\includegraphics[width=.7\columnwidth]{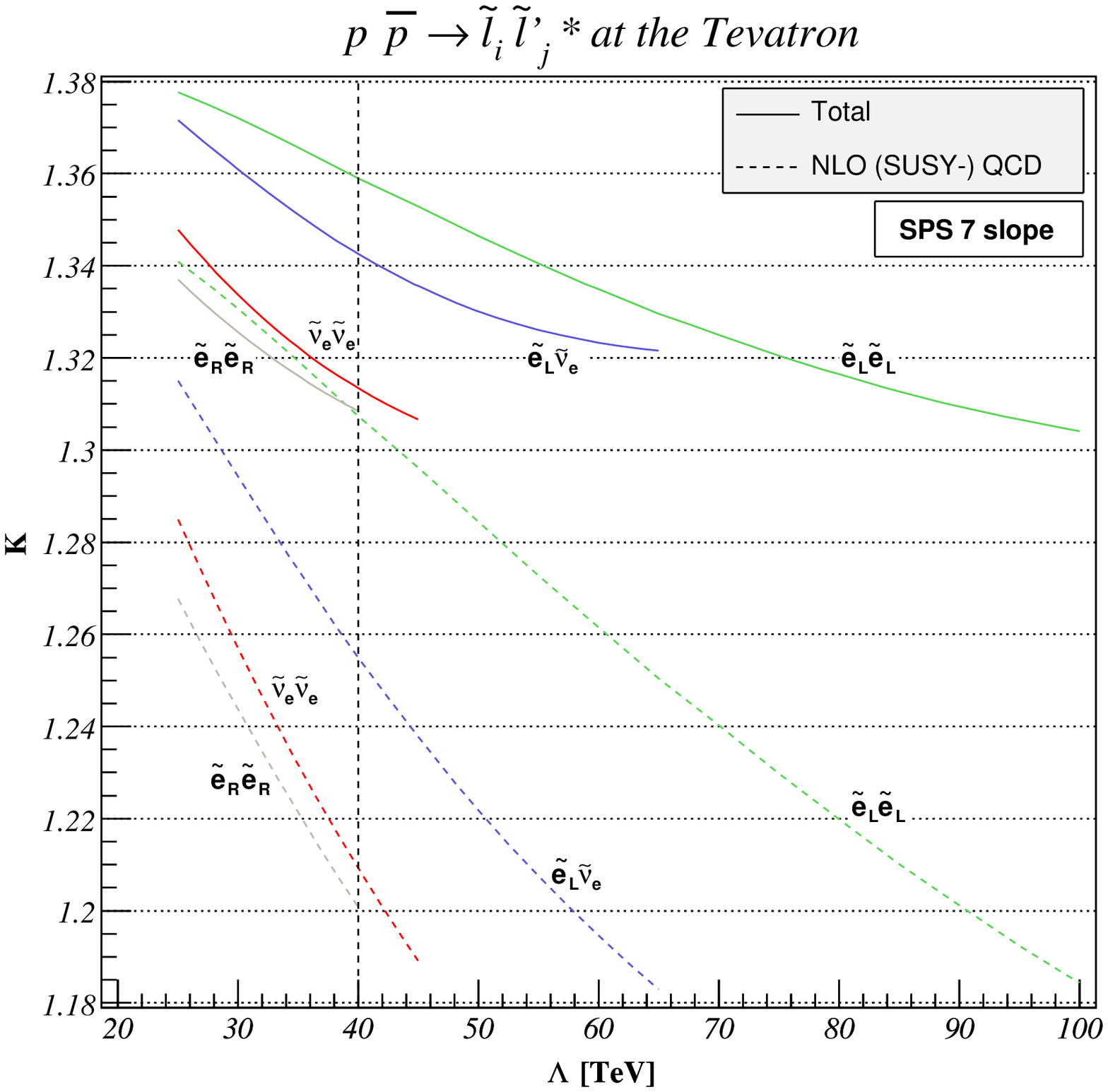}
\caption{\label{fig:8}Total cross sections (top) and $K$-factors as defined
 in Eq.\ (\ref{eq:kglob}) (bottom) for first- (and second-) generation
 slepton-pair and slepton-sneutrino associated production at the Tevatron
 along the model line attached to the SPS 7 benchmark point (vertical dashed
 line). We show the total NLL+NLO matched and the fixed order NLO (SUSY-)QCD
 and LO QCD results.}
\end{figure}
%
In Fig.\ \ref{fig:8}, total cross sections (top) and $K$-factors (bottom)
are shown at the Tevatron, which is expected to produce a total integrated
luminosity of 4--8 fb$^{-1}$. Only left-handed (charged) eigenstates couple
to the weak (electromagnetic) neutral current, so that their respective
cross sections are enhanced. However, even for $\tilde{e}_L$-pair production
the mass-range is limited to masses below 280 GeV, where the cross section
reaches 0.1 fb and at most one event would be produced. NLO and resummation
corrections are clearly important, as they increase the LO prediction by
18 to 28\% (lower part of Fig.\ \ref{fig:8}). At the SPS 7 benchmark point,
the corrections would thus induce a shift in the selectron mass as deduced
from a total cross section measurement by about 8 GeV (cf.\ the two dashed
lines in the upper part of Fig.\ \ref{fig:8}). By comparing the NLO and
total predictions, one observes an increased importance of threshold
resummation for heavier sleptons, as expected.

%
\begin{figure}
\centering
\includegraphics[width=.7\columnwidth]{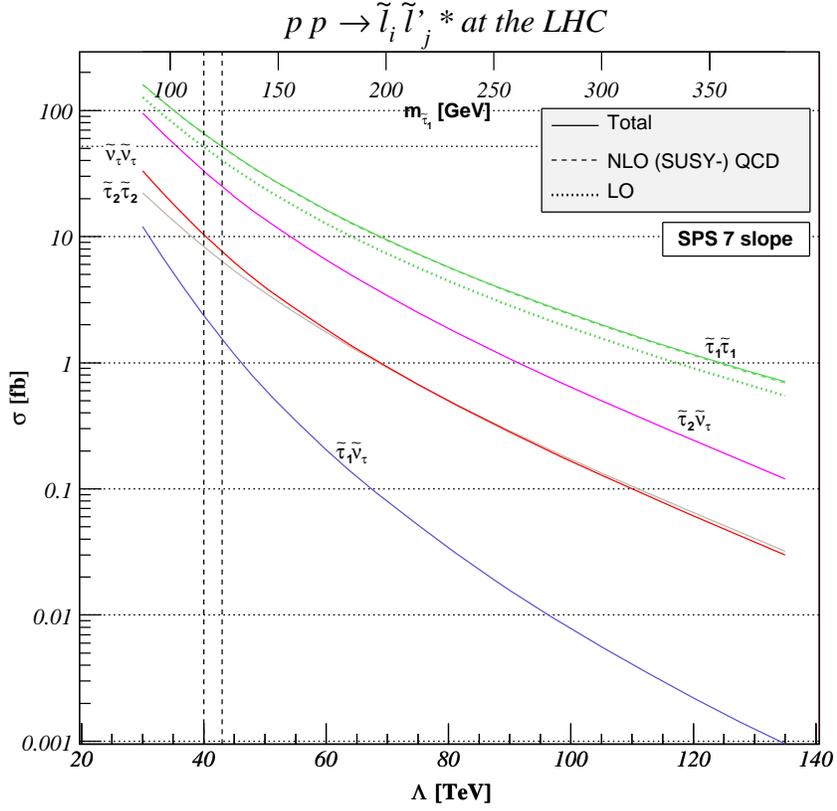}
\includegraphics[width=.7\columnwidth]{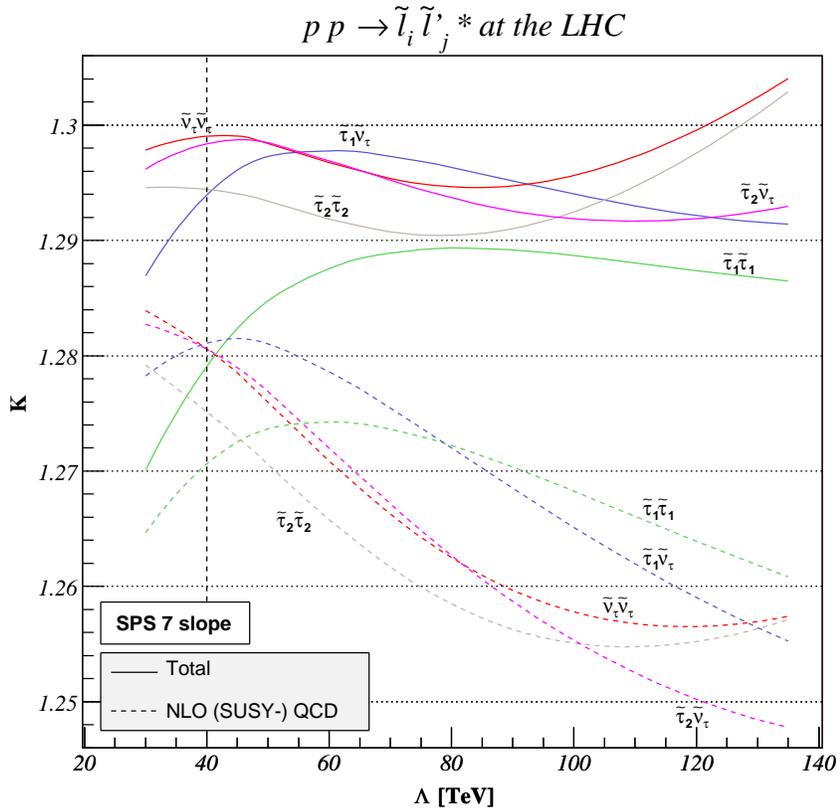}
\caption{\label{fig:9}Total cross sections (top) and $K$-factors as defined
 in Eq.\ (\ref{eq:kglob}) (bottom) for third-generation slepton-pair and
 slepton-sneutrino associated production at the LHC along the model line
 attached to the SPS 7 benchmark point (vertical dashed line). We show the
 total NLL+NLO matched and the fixed order NLO (SUSY-)QCD and LO QCD
 results.}
\end{figure}
%
At the LHC (Fig.\ \ref{fig:9}), even very heavy sleptons can be detected
(see above), but we restrict ourselves to the range $\Lambda\leq135$ TeV.
The NLO and resummed corrections are again large (25--30\%), but the
resummation corrections only become appreciable for large SUSY-breaking
scales (or slepton masses). The largest cross section is obtained for
pair production of the light stau mass eigenstate, even though it has
a large right-handed component. Conversely, the heavier stau mass eigenstate
has a large left-handed component, so that its cross section is less
suppressed. At the SPS 7 benchmark point, the corrections would again induce
a shift in the slepton ($\tilde{\tau}_1$) mass as deduced from a total cross
section measurement by about 8 GeV (cf.\ the two dashed lines in the upper
part of Fig.\ \ref{fig:9}).

%
\begin{figure}
\centering
\includegraphics[width=.9\columnwidth]{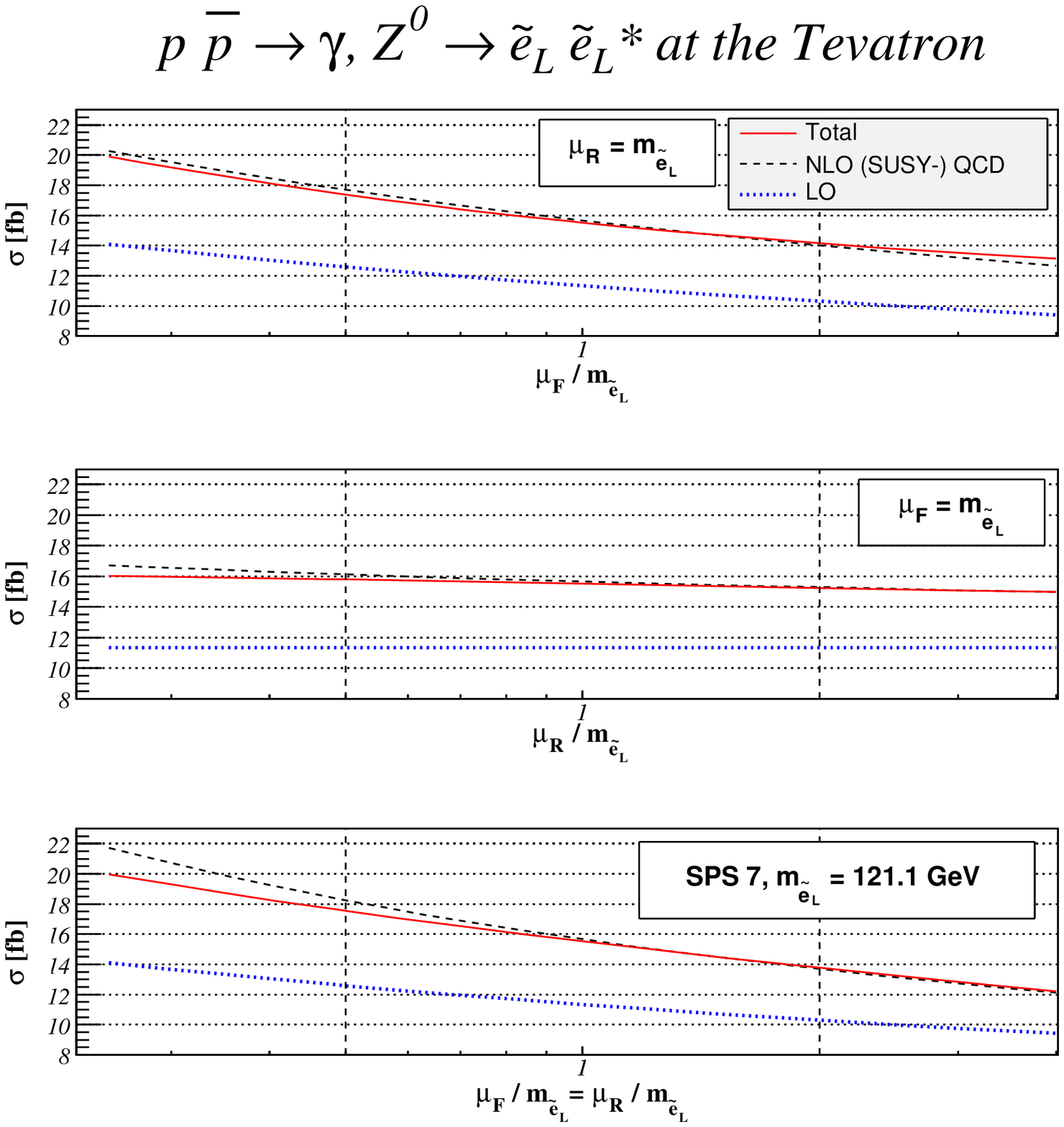}
\caption{\label{fig:10}Dependence of the total cross section for first-
 (and second-) generation slepton pairs at the Tevatron on the factorization
 scale (top), renormalization scale (middle), and both scales (bottom) for
 the SPS 7 benchmark point.}
\end{figure}
%
%
\begin{figure}
\centering
\includegraphics[width=.9\columnwidth]{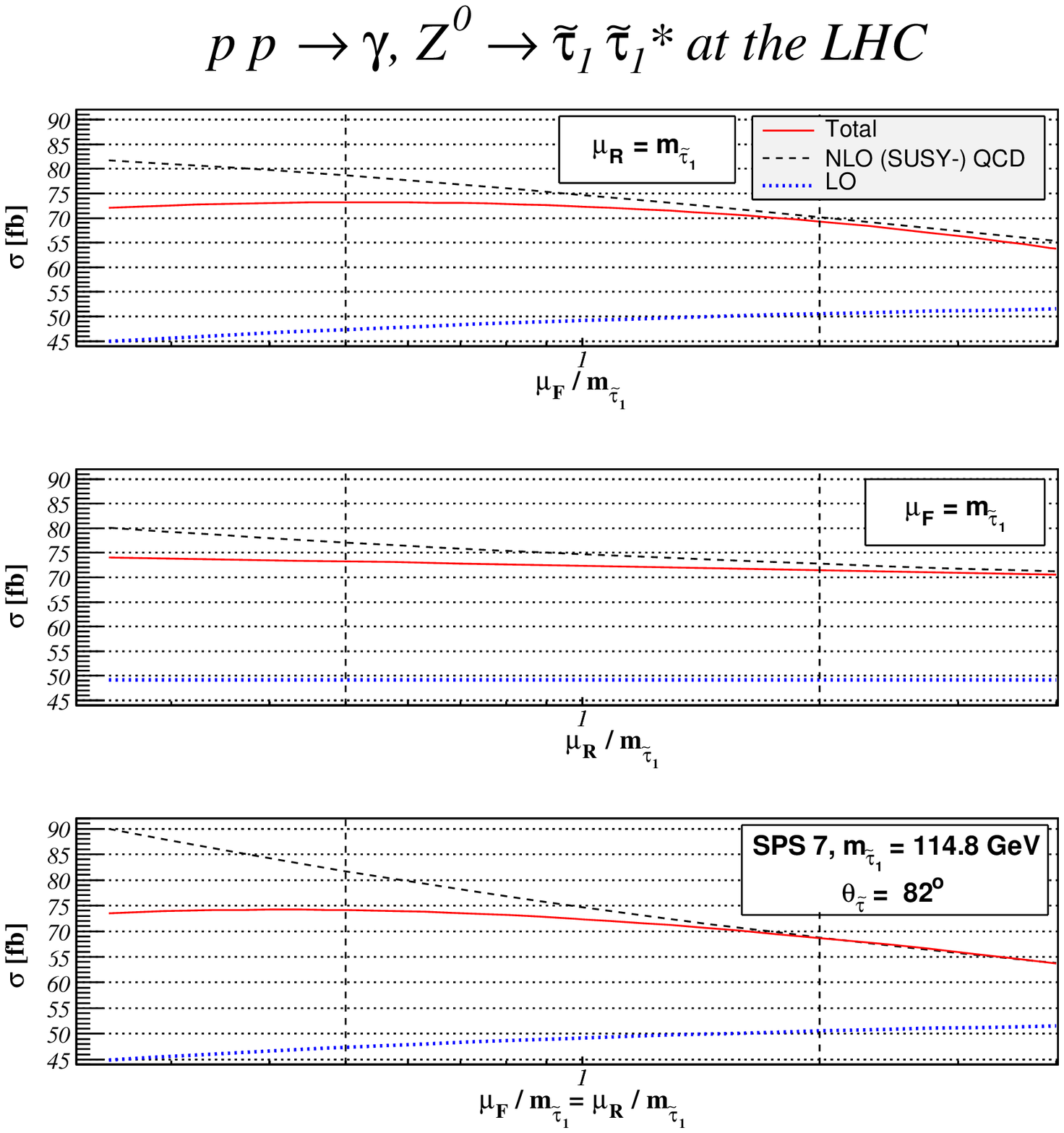}
\caption{\label{fig:11}Dependence of the total cross section for
 third-generation slepton pairs at the LHC on the factorization
 scale (top), renormalization scale (middle), and both scales (bottom) for
 the SPS 7 benchmark point.}
\end{figure}
%
Finally, we consider the theoretical uncertainty of invariant-mass
integrated total cross sections at the Tevatron (Fig.\ \ref{fig:10}) and
the LHC (Fig.\ \ref{fig:11}) as induced by variations of the factorization
scale (top), renormalization scale (middle), or both (bottom). The
$\mu_R$-dependence (middle), which is absent in LO, is first introduced in
NLO, but then tamed by the resummation procedure. On the other hand, the
logarithmic $\mu_F$-dependence (top), already present through the PDFs at
LO, is overcompensated (reduced) at NLO for the LHC (Tevatron) and then
(further) stabilized by resummation. This works considerably better at the
LHC, where at least one quark PDF is sea-like and the PDFs are evaluated at
lower $x$, than at the Tevatron, where both PDFs can be valence-like and are
evaluated at relatively large
$x$. In total, the theoretical uncertainty at the Tevatron (LHC), defined
by the ratio of the cross section difference at $\mu_F=\mu_R=m_{\tilde{l}}/
2$ and $\mu_F=\mu_R=2\, m_{\tilde{l}}$ over their sum, increases
from 20 (7) \% in LO to 29 (17) \% in NLO, but is then reduced again to 23
(8) \% for the resummed-improved prediction.

\section{Conclusions}
\label{sec:5}

In summary, we have presented a first and extensive study on threshold
resummation effects for SUSY-particle production at hadron colliders,
focusing on Drell-Yan like slepton-pair and slepton-sneutrino associated
production in mSUGRA and GMSB scenarios. After confirming the known NLO QCD
corrections and generalizing the NLO SUSY-QCD corrections to the case of
mixing squarks in the virtual loop contributions, we employed the usual
Mellin $N$-space resummation formalism with the minimal prescription for
the inverse Mellin-transform, but improved it by resumming $1/N$-suppressed
and a class of $N$-independent universal contributions. Numerically, our
results increase the theoretical cross sections by 5 to 15\% with respect
to the NLO prediction and stabilize them by reducing the scale dependence
from up to 20\% at NLO to less than 10\% with threshold resummation.

\acknowledgments
We thank M.\ Cacciari for useful discussions concerning the numerical
stability of parton density parameterizations in Mellin $N$-space. This work
was supported by a CNRS/IN2P3 postdoctoral grant and a Ph.D.\ fellowship of
the French ministry for education and research.

\appendix
\section{Sfermion Mixing}
\label{sec:a}

The soft SUSY-breaking terms $A_f$ of the trilinear Higgs-sfermion-sfermion
interaction and the off-diagonal Higgs mass parameter $\mu$ in the MSSM
Lagrangian induce mixings of the left- and right-handed sfermion eigenstates
$\tilde{f}_{L,R}$ of the electroweak interaction into mass eigenstates
$\tilde{f}_{1,2}$. The sfermion mass matrix is given by \cite{Haber:1984rc}
\bea
 {\cal M}^2 &=& \left(\begin{array}{cc} m_{LL}^2+m_f^2 & m_f\,m_{LR}^\ast \\
 m_f\,m_{LR}& m_{RR}^2+m_f^2 \end{array}\right)
\eea
with
\bea
 m_{LL}^2&=&m_{\tilde{F}}^2 + (T_f^3-e_f\,\sin^2\theta_W)\,m_Z^2\,\cos2\beta,\\
 m_{RR}^2&=&m_{\tilde{F}^\prime}^2 + e_f\,\sin^2\theta_W\,m_Z^2\,\cos2\beta,\\
 m_{LR}&=&A_f-\mu^\ast\left\{
 \begin{array}{l}
 \cot\beta\hspace*{3.mm}{\rm for~up-type~sfermions.}
 \\ \tan\beta\hspace*{2.8mm}{\rm for~down-type~sfermions.}
 \end{array}\right.\hspace*{3mm}
\eea
It is diagonalized by a unitary matrix $S^{\tilde{f}}$, $S^{\tilde{f}}\,
{\cal M}^2\,S^{\tilde{f}\dagger}={\rm diag}\,(m_1^2,m_2^2)$, and has the
squared mass eigenvalues
\bea
 m_{1,2}^2=m_f^2+{1\over 2}\Big( m_{LL}^2 + m_{RR}^2
 \mp\sqrt{(m_{LL}^2-m_{RR}^2)^2 + 4\,m_f^2\,|m_{LR}|^2} \Big).
\eea
For real values of $m_{LR}$, the sfermion mixing angle $\theta_{\tilde{f}}$,
$0 \leq \theta_{\tilde{f}} \leq \pi/2$, in
\bea
 S^{\tilde{f}} = \left( \begin{array} {cc}~~\,\cos\theta_{\tilde{f}} &
 \sin\theta_{\tilde{f}} \\
 -\sin\theta_{\tilde{f}} & \cos\theta_{\tilde{f}} \end{array} \right)
 \hspace*{1mm} {\rm with} \hspace*{1mm} \left(
 \begin{array}
 {c} \tilde{f}_1 \\
 \tilde{f}_2
 \end{array} \right) =
 S^{\tilde{f}} \left(
 \begin{array}
 {c} \tilde{f}_L \\ \tilde{f}_R
 \end{array} \right) \hspace*{5mm}&&
\eea
can be obtained from
\begin{equation}
 \tan2\theta_{\tilde{f}}={2\,m_f\,m_{LR}\over m_{LL}^2-m_{RR}^2}.
\end{equation}
If $m_{LR}$ is complex, one may first choose a suitable phase rotation
$\tilde{f}_R^\prime = e^{i\phi}\tilde{f}_R$ to make the mass matrix real and
then diagonalize it for $\tilde{f}_L$ and $\tilde{f}_R^\prime$. $\tan
\beta=v_u/v_d$ is the (real) ratio of the vacuum expectation values of the
two Higgs fields, which couple to the up-type and down-type (s)fermions. The
soft SUSY-breaking mass terms for left- and right-handed sfermions are
$m_{\tilde{F}}$ and $m_{\tilde{F}^\prime}$ respectively.

\section{SUSY-QCD Form Factors for Mixed Squark Mass Eigenstates}
\label{sec:b}

At next-to-leading order in perturbative QCD, SUSY-QCD form factors are
induced by one-loop diagrams involving gluinos and (generally mixed) squark
mass eigenstates. Those appearing in the neutral-current cross section of
Eq.\ (\ref{eq:nenlo}) are given by
\bea
 f_\gamma &=& 2 + \sum_{i=1,2} \Bigg[\frac{2\, m_{\tilde{g}}^2 - 2\,
 m_{\tilde{q}_i}^2 + M^2}{M^2}\,
 \Big(B_{0f}\left(M^2, m_{\tilde{q}_i}^2, m_{\tilde{q}_i}^2\right)
 - B_{0f} \left(0, m_{\tilde{g}}^2, m_{\tilde{q}_i}^2\right)\Big)
 +\left(m_{\tilde{q}_i}^2 - m_{\tilde{g}}^2\right)\,
 B^\prime_{0f}\left(0, m_{\tilde{g}}^2,
 m_{\tilde{q}_i}^2\right) \nonumber \\
 &+&2 \, \frac{m_{\tilde{g}}^4 + (M^2-2\, m_{\tilde{q}_i}^2)\,
 m_{\tilde{g}}^2 + m_{\tilde{q}_i}^4}{M^2}\, C_{0f}\left(0, M^2, 0,
 m_{\tilde{q}_i}^2, m_{\tilde{g}}^2, m_{\tilde{q}_i}^2\right)
 \Bigg],\\
 f_{\gamma Z} &=& 2\, (L_{Z q q} + R_{Z q q}) \nonumber\\
 &+&\sum_{i=1,2}\Bigg[2\,\frac{\left(2\, m_{\tilde{g}}^2 - 2\,
 m_{\tilde{q}_i}^2 + M^2\right)\,{\rm Re}\Big[L_{Z \tilde{q}_i
 \tilde{q}_i} + R_{Z \tilde{q}_i
 \tilde{q}_i}\Big]}{M^2}\,B_{0f}\left(M^2,
 m_{\tilde{q}_i}^2, m_{\tilde{q}_i}^2\right) \Bigg] \nonumber\\
 &-&\sum_{i=1,2}\Bigg[2\,\frac{\left(2\, m_{\tilde{g}}^2 - 2\,
 m_{\tilde{q}_i}^2 + M^2\right)\,\Big(L_{Z q q}\,
 \Big|S^{\tilde{q}}_{i1}\Big|^2  + R_{Z q q}\,
 \Big|S^{\tilde{q}}_{i1}\Big|^2 \Big)}{M^2}\, B_{0f}\left(0,
 m_{\tilde{g}}^2, m_{\tilde{q}_i}^2\right)\Bigg]  \nonumber\\
 &+& \sum_{i=1,2}\Bigg[\left(m_{\tilde{q}_i}^2 - m_{\tilde{g}}^2
 \right) (L_{Z q q}+R_{Z q q}) \, B_{0f}^\prime\left(0,
 m_{\tilde{g}}^2,
 m_{\tilde{q}_i}^2\right)\Bigg]\nonumber\\
 &+& \sum_{i=1,2}\Bigg[ 4 \, \frac{\left(m_{\tilde{g}}^4 + (M^2 -
 2\, m_{\tilde{q}_i}^2)\, m_{\tilde{g}}^2 +
 m_{\tilde{q}_i}^4\right)\,{\rm Re}\Big[L_{Z \tilde{q}_i
 \tilde{q}_i} + R_{Z \tilde{q}_i \tilde{q}_i}\Big]}{M^2}\,
 C_{0f}\left(0, M^2, 0, m_{\tilde{q}_i}^2, m_{\tilde{g}}^2,
 m_{\tilde{q}_i}^2\right)\Bigg],~~{\rm and}\\
 f_Z &=& 2\, \left(L_{Z q q}^2+R_{Z q q}^2\right) \nonumber \\
 &+& \sum_{i,j=1,2} \left[2\, \frac{\left(2\,m_{\tilde{g}}^2 -
 m_{\tilde{q}_i}^2 - m_{\tilde{q}_j}^2 +
 M^2\right)}{M^2}\,\Big|L_{Z \tilde{q}_i \tilde{q}_j} + R_{Z
 \tilde{q}_i \tilde{q}_j}\Big|^2\, B_{0f}\left(M^2,
 m_{\tilde{q}_i}^2, m_{\tilde{q}_j}^2\right)
 \right]\nonumber \\
 &-&\sum_{i=1,2}\Bigg[2\, \frac{\left(2\, m_{\tilde{g}}^2 - 2\,
 m_{\tilde{q}_i}^2 + M^2\right)\,\Big(L^2_{Z q
 q}\,\Big|S^{\tilde{q}}_{i1}\Big|^2 + R^2_{Z q q}\,
 \Big|S^{\tilde{q}}_{i2}\Big|^2 \Big)}{M^2}\,B_{0f}\left(0,
 m_{\tilde{g}}^2, m_{\tilde{q}_i}^2\right) \Bigg] \nonumber \\
 &+& \sum_{i=1,2}\Bigg[\left(m_{\tilde{q}_i}^2 -
 m_{\tilde{g}}^2\right) \left(L_{Z q q}^2 + R_{Z q q}^2\right)\,
 B_{0f}^\prime\left(0, m_{\tilde{g}}^2,
 m_{\tilde{q}_i}^2\right)\Bigg]\nonumber\\
 &+&\sum_{i,j=1,2}\left[4\,\frac{\left(m_{\tilde{g}}^4 + \left(M^2 -
 m_{\tilde{q}_i}^2 - m_{\tilde{q}_j}^2\right)\, m_{\tilde{g}}^2 +
 m_{\tilde{q}_i}^2\, m_{\tilde{q}_j}^2 \right)}{M^2}\, \Big|L_{Z
 \tilde{q}_i \tilde{q}_j} + R_{Z \tilde{q}_i \tilde{q}_j}\Big|^2\,
 C_{0f}\left(0, M^2, 0, m_{\tilde{q}_i}^2, m_{\tilde{g}}^2,
 m_{\tilde{q}_j}^2\right) \right],\hspace*{5mm}
\eea
and the one appearing in the charged-current cross section of Eq.\
(\ref{eq:chnlo}) by
\bea
f_W &=& 2\,\Big|L_{W q q^\prime}\Big|^2 \nonumber \\
&+& \sum_{i,j=1,2} \Bigg[ \frac{2 \left(2\, m_{\tilde{g}}^2 -
m_{\tilde{q}_i}^2 - m_{\tilde{q}^\prime_j}^2 + M^2\right)
\Big|L_{W \tilde{q}_i \tilde{q}_j^\prime}\Big|^2}{M^2}\,
B_{0f}\left(M^2, m_{\tilde{q}_i}^2,
m_{\tilde{q}^\prime_j}^2\right)\Bigg] \nonumber \\
&-& \sum_{\tilde{Q}=\tilde{q},\tilde{q}^\prime} \, \sum_{i=1,2}
\Bigg[\frac{\left(2\, m_{\tilde{g}}^2 - 2\, m_{\tilde{Q}_i}^2 +
M^2\right) B_{0f}\left(0, m_{\tilde{g}}^2, m_{\tilde{Q}_i}^2
\right)\Big|L_{W q q^\prime} \, S^{\tilde{Q}}_{i1}\, \Big|^2
}{M^2} \Bigg]\nonumber
\\
&+& \sum_{\tilde{Q}=\tilde{q},\tilde{q}^\prime} \, \sum_{i=1,2}
\Bigg[\frac{1}{2} \left(m_{\tilde{Q}_i}^2 -
m_{\tilde{g}}^2\right)\, \Big|L_{W q q^\prime}\Big|^2\,
B_{0f}^\prime\left(0, m_{\tilde{g}}^2, m_{\tilde{Q}_i}^2\right)
\Bigg]\nonumber
\\ &+& \sum_{i,j=1,2}\Bigg[\frac{4 \left(m_{\tilde{g}}^4 -
\left(m_{\tilde{q}_i}^2 + m_{\tilde{q}^\prime_j}^2 - M^2\right)\,
m_{\tilde{g}}^2 + m_{\tilde{q}_i}^2\,
m_{\tilde{q}^\prime_j}^2\right) \Big|L_{W \tilde{q}_i
\tilde{q}_j^\prime}\Big|^2}{M^2}\, C_{0f}\left(0, M^2, 0,
m_{\tilde{q}^\prime_i}^2, m_{\tilde{g}}^2,
m_{\tilde{q}_j}^2\right)\Bigg].
\eea
The functions $B_{0f}(p^2, m_1^2, m_2^2)$, $B_{0f}^\prime(p^2,m_1^2,m_2^2)$
and $C_{0f}(p_1^2, (p_1 + p_2)^2, p_2^2, m_1^2,m_2^2, m_3^2)$ are the finite
parts of the scalar two- and three-point functions
\bea
 B_0(p^2, m_1^2,m_2^2) &=& \mu_R^{2\varepsilon}\, \int \frac{{\rm d}^Dq}
 {i \pi^2}\,\frac{1}{(q^2 - m_1^2)\, ((q+p)^2 -m_2^2)}~,~\\
 B_0^\prime(p^2, m_1^2, m_2^2) &=& \frac{{\rm d}B_0(k^2,
 m_1^2, m_2^2)}{{\rm d}k^2}\Big|_{k^2=p^2}~,~~{\rm and}\\
 C_0(p_1^2, (p_1 + p_2)^2, p_2^2, m_1^2, m_2^2, m_3^2) &=&
 \mu_R^{2\varepsilon}\, \int \frac{{\rm d}^Dq}{i \pi^2}\,
 \frac{1}{(q^2 - m_1^2)\, ((q+p_1)^2 - m_2^2)\, ((q+p_1+p_2)^2 - m_3^2)}.
\eea
Our results agree with those of Ref.\ \cite{Djouadi:1999ht} in the case of
mass-degenerate non-mixing squarks.


\end{document}